\documentclass[manuscript]{aastex631}

\definecolor{Red}{rgb}{1,0,0}

\usepackage{bm}

\usepackage{amssymb}
\usepackage{lipsum}
\usepackage{amsmath,amsfonts}
\usepackage{rotating}
\usepackage{longtable}

\received{} %June 1, 2019}
\revised{} %January 10, 2019}
\accepted{} %\today}
\submitjournal{The Astrophysical Journal Supplement Series}

\shorttitle{Identification of Ion-Kinetic Instabilities with Machine Learning}
\shortauthors{Sadykov et al.}
\begin{document}

\title{Identification of Ion-Kinetic Instabilities in Hybrid-PIC Simulations of Solar Wind Plasma with Machine Learning}

\author[0000-0002-4001-1295]{Viacheslav M Sadykov}
\affiliation{Physics \& Astronomy Department, 
Georgia State University, 
Atlanta, GA 30303, USA}

\author[0000-0003-0602-6693]{Leon Ofman}
\affiliation{The Catholic University of America, Washington, DC 20064, USA}
\affiliation{Heliophysics Science Division, NASA Goddard Space Flight Center, 
Greenbelt, MD 20771, USA}
\affiliation{Visiting, Tel Aviv University, Tel Aviv, Israel}

\author[0000-0002-5240-044X]{Scott A Boardsen}
\affiliation{Goddard Planetary Heliophysics Institute, University of Maryland, Baltimore, MD 21250, USA}
\affiliation{Heliophysics Science Division, NASA Goddard Space Flight Center, 
Greenbelt, MD 20771, USA}

\author[0000-0001-6018-9018]{Yogesh}
\affiliation{The Catholic University of America, Washington, DC 20064, USA}
\affiliation{Heliophysics Science Division, NASA Goddard Space Flight Center, 
Greenbelt, MD 20771, USA}

\author[0000-0002-3808-3580]{Parisa Mostafavi}
\affiliation{Johns Hopkins Applied Physics Laboratory, 
Laurel, MD 20723, USA}

\author[0000-0002-6849-5527]{Lan K Jian}
\affiliation{Heliophysics Science Division, 
NASA Goddard Space Flight Center, 
Greenbelt, MD 20771, USA}

\author[0000-0001-6038-1923]{Kristopher Klein}
\affiliation{Lunar and Planetary Laboratory, University of Arizona, Tucson, AZ 85721, USA}

\author[0000-0002-7365-0472]{Mihailo Martinovi\'{c}}
\affiliation{Lunar and Planetary Laboratory, University of Arizona, Tucson, AZ 85721, USA}

\begin{abstract}
Analysis of ion-kinetic instabilities in solar wind plasmas is crucial for understanding energetics and dynamics throughout the heliosphere, as evident from spacecraft observations of complex ion velocity distribution functions (VDFs) and ubiquitous ion-scale kinetic waves. 
In this work, we explore machine learning (ML) and deep learning (DL) classification models to identify unstable cases of ion VDFs driving kinetic waves.
Using 34 hybrid particle-in-cell simulations of kinetic protons and $\alpha$-particles initialized using plasma parameters derived from solar wind observations, we prepare a dataset of nearly 1600 VDFs representing stable/unstable cases and associated plasma and wave properties. 
We compare feature-based classifiers applied to VDF moments, such as Support Vector Machine and Random Forest, with DL convolutional neural networks (CNN) applied directly to VDFs as images in the gyrotropic velocity plane.
The best-performing classifier, Random Forest, has an accuracy of $0.96\pm0.01$, and a true skill score of $0.89\pm0.03$, with the majority of missed predictions made near stability thresholds.
We study how the variations of the temporal derivative thresholds of anisotropies and magnetic energies and sampling strategies for simulation runs affect classification.
CNN-based models have the highest accuracy of $0.88\pm0.18$ among all considered if evaluated on the runs entirely not used during the model training. 
The addition of the $E_{\perp}$ power spectrum as an input for the ML models leads to the improvement of instability analysis for some cases. 
The results demonstrate the potential of ML and DL for the detection of ion-scale kinetic instabilities using spacecraft observations of solar wind and magnetospheric plasmas.
\end{abstract}

\keywords{Sun, Plasma stability, Plasma waves, Classification algorithms, Convolutional neural networks}

\section{Introduction}

The solar wind (SW) is a continuous stream of charged particles (primarily electrons, protons, and $\alpha$-particles) directed away from the Sun and largely organized by the interplanetary magnetic field. 
Because of departures from local thermodynamic equilibrium (LTE), the presence of a preferred direction dictated by the background large-scale magnetic field $\mathbf{B}_0$ and its nearly collisionless nature, the particle populations in the solar wind are often observed to be be non-Maxwellian, requiring kinetic treatment, i.e., the description of the evolution of their Velocity Distribution Functions (VDFs) subject to wave-particle interactions and electro-magnetic forces, such as with Vlasov's equations. 
Since the VDFs of solar wind ions can be non-Maxwellian and experience temperature anisotropy (the state when the temperature along the external magnetic field, $T_{||}$, is not equal to temperature across the magnetic field, $T_{\perp}$, in particular encountered in the inner heliosphere \citep{Marsch2012helios}). 
The VDFs also can consist of several distinct components (core and beam populations) or show the relative velocity drifts of particle species \citep[for example, $p$-$\alpha$ drifts,][]{Alterman:2018,Mostafavi2022alpha, Johnson_2024}. 
In these scenarios, there is free energy stored in the plasma which can support the development of kinetic instabilities for certain ranges of plasma parameters \citep{Gary:1993}.
The development of anisotropy-driven \citep[such as ion-cyclotron, mirror, parallel and oblique firehose,][]{Verscharen_etal_2013_AIC_instability,Verscharen2016collisionless} or relative drift-driven \citep{Kasper2013PhRvL.110i1102K} instabilities redistributes this free energy into ion-scale electromagnetic waves which, in turn, interact with the particles and modify ion VDFs. 
Such collisionless wave-particle interaction processes are important in the complex and highly dynamic solar wind plasma as it evolves throughout the heliosphere, as observed in spacecraft data. 
While complex electron VDFs can drive instabilities \citep{Verscharen:2022b}, the focus of the present study is on protons and $\alpha$ particles as these ions carry the bulk of the solar wind kinetic energy and mass flux.

The launch of the Parker Solar Probe (PSP) \citep{Fox2016SSRv..204....7F} provided new insights into the properties of a very young solar wind observed at distances as close as  $\sim$10 solar radii ($R_\odot$) from the Sun, and confirmed the ubiquitous presence of ion-scale kinetic waves \citep{Raouafi2023parker}. 
The measurements of the $p$ and $\alpha$ particle populations by the Solar Probe ANalyzer for Ions (SPAN-I) instrument \citep{Livi2022spani} revealed the presence of the ion particle beams and anisotropic VDFs accompanied with the ion-scale wave activity \citep{Verniero2020parker,Ofman2022ApJ...926..185O,Ofman2023ApJ...954..109O,McManus:2024,Klein2021ApJ...909....7K,Shankarappa2024ApJ...973...20S}, pointing to the presence of ion-kinetic instability processes in progress. 
Overall, PSP provides a window into how the kinetic instabilities develop in the young solar wind, as well as how they would energize the solar wind at further distances from the Sun.

Identification of the unstable plasma conditions in the solar wind based on the observations of VDFs is a challenging problem. 
In particular, the plasma `parcels' in the solar wind are rarely observed in more than one point \citep[for the development of multi-spacecraft tracking strategies, see, for example,][]{Berriot2024identification}. 
Therefore, the VDFs for the given parcel are sampled only `once' in the spacecraft frame at any given time as they are integrated in a short time interval in a single point measurement, and the VDF evolution in the solar wind frame is not directly available from such observations. 
In such cases, the common approach is to perform a linear stability analysis of the observed VDFs by solving the linearized Vlasov's equation  \citep{Verscharen2018alps,Klein2015predicted} with one of many simplifying assumptions that no significant wave-particle interaction developed in the SW plasmas at the given moment of observations. 
Given that there is limited observational capability to `track' the plasma parcel in the solar wind moving frame while the instability develops, analytical intractability and the numerically difficulty of solving the full multi-dimensional nonlinear Vlasov's equations \citep[e.g.][]{Pal18}, physics-based simulations such as PIC or hybrid-PIC represent the main viable approach to study the non-linear temporal evolution of the kinetic instabilities. 
In particular, the hybrid-PIC simulations \citep{Ofman2010wave} can allow one to model the ion-scale wave processes and dynamics efficiently while preserving the main essential physics. 
If initialized with linearly unstable VDFs, the simulations will eventually evolve into a regime where waves and ion distributions interact non-linearly, providing the possibility of indirectly studying such regimes in the solar wind.

The identification of the unstable ion VDF distributions could be, in principle, formulated as a binary classification problem for machine learning (ML). 
The time moments of the spatially-integrated VDF evolutions where there is a significant growth of the wave activity and the related changes in the VDF itself could be marked as kinetically unstable (and represent unstable cases) and, correspondingly, the periods with no wave activity and unchanged VDFs could be marked as stable (and represent stable cases).
Such identification and categorization of the ion VDF properties with respect to the linear stability criteria has been done previously in \cite{Martinovic2023ApJ...952...14M}. 
Using the Plasma in a Linear Uniform Magnetized Environment (PLUME) linear dispersion solver \citep{Klein2015predicted} combined with the Nyquist instability criterion \citep{Klein:2017}, the authors generated a training set of linearly stable and unstable VDFs prescribed with the temperature anisotropy $A=T_{\perp}/T_{\parallel}$ and plasma $\beta{}_{\parallel}$ (the ratio of thermal to magnetic energy, $8\pi{}nk_BT_{\parallel}/B^{2}$) parameters for the $p$ core and beam, as well as $\alpha$ particle distributions, and the relative densities of the populations and velocity drifts of the components. 
The overall performance of the developed classifier for instability detection had a superior accuracy of $\sim{}0.96-0.99$, depending on the number of ion species involved. 
The approach involved testing the stability of distributions with respect to the linear stability analysis criteria, and therefore, the consideration of the non-linear effects remains of interest, which is in the scope of the current work.

In this work, we present an effort to develop ML-driven models for identifying instabilities in the solar wind based on the computed set of nonlinear hybrid-PIC simulation runs that consider fully self-consistently the wave-particle interactions for the ions. 
These nonlinear interactions may affect the SW plasma stability considerably, in comparison to linear stability analysis, therefore, we extend the previous work into the nonlinear regime.
Section~\ref{sec:hybridpic} describes the hybrid-PIC simulation runs used in this work. 
Section~\ref{sec:MLmodel} provides the description of the considered machine learning models, their optimization, and evaluation strategies. 
Section~\ref{sec:results} summarizes the results of this work, and is followed by the discussion and summary presented in Section~\ref{sec:discussion}.

\section{Instabilities in Hybrid-PIC Simulations}\label{sec:hybridpic}

\subsection{Processing of Hybrid-PIC simulation runs}

We leverage existing hybrid-PIC runs of the solar wind plasma \citep[see, e.g.,][]{Ofman2010wave}, and simulate new cases using PSP-inspired SW parameters at perihelia. 
In these simulations, protons and $\alpha$ particles are modeled explicitly as super-particles in collisionless approximation (i.e., there are no ion-ion Coulomb interactions involved), while electrons represent the charge-neutralizing fluid \citep[e.g.,][]{WO93}. 
Effectively, this approximation averages over the smaller spatiotemporal scales of electron kinetic motions by using the generalized Ohm's law, while allowing one to resolve the larger (compared to electrons) ion-scale dynamics and field fluctuations. 
Some hybrid-PIC simulations corresponding to the plasma conditions observed in the young solar wind by PSP have been applied recently in \cite{Ofman2022ApJ...926..185O,Ofman2023ApJ...954..109O}. 
Additionally, we utilize the simulations recently used in understanding the hot anisotropic ion beams in solar wind \citep{Ofman2025PIC}, motivated by PSP observations, and the investigation of the interaction of ion-scale waves with proton velocity distributions \citep{Yogesh2025PIC}.

The initialization parameters of 34 hybrid-PIC simulation runs considered in this work, including $p$ and $\alpha$ core and beam populations, when applicable, are presented in Table~\ref{table:hybrid-pic-runs}. 
The table summarizes the number of species and components used in the simulations, as well as some of their physical properties for the initial time of the simulation (the velocities of the components in the Alfv\'{e}n speed, $V_{A}=B/\sqrt{4 \pi (n_p m_p + n_\alpha m_\alpha)}$, units, relative population abundances, anisotropies, and $\beta{}_{\parallel}$. 
Please note that these are not necessarily the core-only $p$ and $\alpha$ populations: 8 simulations included the core and the beam components separately for both $p$ and $\alpha$ particles; 13 simulation runs included only the $p$ core and beam components, with no $\alpha$ particles involved. 
To capture the instabilities driven by the relative $p$-$\alpha$ drifts \citep{Kasper2013PhRvL.110i1102K,McManus:2024}, some simulation runs were initialized with these drifts: for example, simulation \#8 features a relative drift between the species of 2$V_{A}$. 
Simulations with both sub-Alfv\'{e}nic and super-Alfv\'{e}nic drifts are also represented in the dataset. 
Additionally, we simulate both the `high' (10\% by number density) and `depleted' ($\sim$1.4\% by number density) abundance of $\alpha$ particles, where both of these possibilities are encountered in the solar wind observations \citep{Ofman2024ApJ...970L..16O}, as well as varying plasma $\beta_{\parallel}$ for the species.

\begin{sidewaystable}[htbp]
\caption{Initialization parameters for hybrid-PIC simulations considered in this work. 
Commas separate the parameters for the different species and core/beam populations.}
\begin{center}
\tiny
\begin{tabular}{cccccc}
\hline
\textbf{Simulation run \#} &   \textbf{Species}   &   \textbf{Velocities, $V_{||}/V_{A}$} &   \textbf{Anisotropies, $T_{\perp}/T_{||}$}  &   \textbf{Relative Populations}  &   \textbf{$\beta{}_{||}$}  \\
\hline
1 & (2) p core, $\alpha$ core & 0.00, 0.00 & 1.0, 0.5 & 0.9, 0.05 & 0.5, 0.5 \\ 
2 & (2) p core, $\alpha$ core & 0.00, 0.00 & 1.0, 0.75 & 0.9, 0.05 & 1.0, 1.0 \\
3 & (2) p core, $\alpha$ core & -0.01, 0.41 & 3.36, 2.0488 & 0.986, 0.007 & 0.08, 0.024 \\ 
4 & (2) p core, $\alpha$ core & 0.00, 0.00 & 4.3, 1.6 & 0.986, 0.007 & 0.08, 0.024 \\ 
5 & (2) p core, $\alpha$ core & 0.00, 0.32 & 4.3, 1.6 & 0.986, 0.007 & 0.08, 0.024 \\ 
6 & (2) p core, $\alpha$ core & 0.00, 0.32 & 1.86, 1.0 & 0.986, 0.007 & 0.10, 0.028 \\ 
7 & (2) p core, $\alpha$ core & 0.00, 0.00 & 1.0, 10.0 & 0.9, 0.05 & 0.0413, 0.0413 \\ 
8 & (2) p core, $\alpha$ core & -0.20, 1.80 & 1.0, 1.0 & 0.9, 0.05 & 0.0413, 0.0413 \\ 
9 & (2) p core, $\alpha$ core & -0.15, 1.35 & 1.0, 1.0 & 0.9, 0.05 & 0.0413, 0.0413 \\ 
10 & (2) p core, p beam & -0.15, 1.35 & , 1.0, 1.0 & 0.9, 0.10 & 0.0413, 0.010325 \\ 
11 & (2) p core, p beam & -0.10, 0.90 & 2.0, 2.0 & 0.90, 0.10 & 0.214, 0.858 \\ 
12 & (3) p core, p beam, $\alpha$ core & -0.20, 1.80, 0.00 & 1.0, 1.0, 1.0 & 0.819, 0.091, 0.045 & 0.429, 0.0916, 0.429 \\ 
13 & (2) p core, p beam & -0.20, 1.80 & 1.0, 1.0 & 0.90, 0.10 & 0.429, 0.858 \\ 
14 & (2) p core, p beam & -0.20, 1.80 & 1.0, 2.0 & 0.90, 0.10 & 0.429, 0.858 \\ 
15 & (2) p core, p beam & -0.20, 1.80 & 2.0, 2.0 & 0.90, 0.10 & 0.214, 0.858 \\ 
16 & (4) p core, p beam, $\alpha$ core, $\alpha$ beam & -0.20, 1.80, -0.29, 2.29 & 1.0, 1.0, 1.0, 1.0 & 0.819, 0.091, 0.04, 0.005 & 0.429, 0.0916, 0.429, 0.0916 \\ 
17 & (4) p core, p beam, $\alpha$ core, $\alpha$ beam & -0.20, 1.80, -0.67, 1.33 & 1.0, 1.0, 1.0, 1.0 & 0.819, 0.091, 0.03, 0.015 & 0.429, 0.0916, 0.429, 0.0916 \\ 
18 & (2) p core, $\alpha$ core & -0.12, 1.08 & 1.0, 0.75 & 0.9, 0.05 & 0.10, 0.028 \\ 
19 & (2) p core, $\alpha$ core & -0.02, 1.18 & 3.35, 2.05 & 0.986, 0.007 & 0.10, 0.028 \\ 
20 & (2) p core, $\alpha$ core & -0.02, 1.48 & 1.5, 1.0 & 0.986, 0.007 & 0.10, 0.028 \\ 
21 & (2) p core, p beam & -0.23, 0.74 & 1.349, 0.503 & 0.765, 0.235 & 0.060, 0.398 \\ 
22 & (2) p core, p beam & -0.43, 0.24 & 2.26, 0.420 & 0.362, 0.638 & 0.044, 0.540 \\ 
23 & (2) p core, p beam & -0.34, 0.48 & 1.542, 0.671 & 0.587, 0.413 & 0.057, 0.354 \\ 
24 & (2) p core, p beam & -0.34, 0.55 & 1.764, 0.620 & 0.616, 0.384 & 0.054, 0.382 \\ 
25 & (2) p core, p beam & -0.34, 0.48 & 1.54, 0.459 & 0.584, 0.416 & 0.046, 0.299 \\ 
26 & (2) p core, p beam & -0.41, 0.29 & 1.407, 0.380 & 0.410, 0.590 & 0.047, 0.510 \\ 
27 & (2) p core, p beam & -0.32, 0.62 & 1.406, 0.424 & 0.661, 0.339 & 0.059, 0.447 \\ 
28 & (4) p core, p beam, $\alpha$ core, $\alpha$ beam & -0.14, 1.26, -0.22, 1.78 & 2.0, 2.0, 2.0, 2.0 & 0.819, 0.091, 0.040, 0.005 & 0.429, 0.429, 0.429, 0.429 \\ 
29 & (2) p core, p beam & -0.20, 1.80 & 2.0, 2.0 & 0.90, 0.10 & 0.429, 0.429 \\ 
30 & (4) p core, p beam, $\alpha$ core, $\alpha$ beam & -0.20, 1.80, -0.16, 1.28 & 1.0, 1.0, 1.0, 1.0 & 0.819, 0.091, 0.040, 0.005 & 0.214, 0.858, 0.214, 0.858 \\ 
31 & (4) p core, p beam, $\alpha$ core, $\alpha$ beam & -0.20, 1.80, -0.16, 1.28 & 2.0, 2.0, 2.0, 2.0 & 0.819, 0.091, 0.040, 0.005 & 0.214, 0.858, 0.214, 0.858 \\ 
32 & (4) p core, p beam, $\alpha$ core, $\alpha$ beam & -0.20, 1.80, -0.16, 1.28 & 2.0, 2.0, 2.0, 2.0 & 0.819, 0.091, 0.040, 0.005 & 0.429, 0.429, 0.429, 0.429 \\ 
33 & (4) p core, p beam, $\alpha$ core, $\alpha$ beam & -0.20, 1.80, -0.22, 1.78 & 1.0, 1.0, 1.0, 1.0 & 0.819, 0.091, 0.040, 0.005 & 0.429, 0.429, 0.429, 0.429 \\ 
34 & (4) p core, p beam, $\alpha$ core, $\alpha$ beam & -0.20, 1.80, -0.22, 1.78 & 2.0, 2.0, 2.0, 2.0 & 0.819, 0.091, 0.040, 0.005 & 0.429, 0.429, 0.429, 0.429 \\
\hline
\end{tabular}
\label{table:hybrid-pic-runs}
\end{center}
\end{sidewaystable}

Depending on how fast the evolution of the ion particle distributions happens, each simulation was run for a different amount of time, measured in units of inverse proton gyrofrequency, $\Omega_{p}^{-1}$, and had a different sampling frequency of the sampled snapshots depending on the expected evolution rate. 
For example, simulation run \#3 was run for 600\,$\Omega_{p}^{-1}$ and sampled every 12.5\,$\Omega_{p}^{-1}$ based on its expected fast evolution because of an instability development \citep{Ofman2023ApJ...954..109O}. 
In contrast, simulation run \#20, with a more slowly-developing instability, lasted 4000\,$\Omega_{p}^{-1}$ and was sampled every 125\,$\Omega_{p}^{-1}$. 
The total number of VDF samples from all 34 runs utilized in the present study was 1596. 
The large number of VDFs used was aimed at improving the validity of the ML analysis.

\begin{figure}[htbp]
\begin{minipage}{1.00\linewidth}
    \centering
    \includegraphics[width=0.90\linewidth]{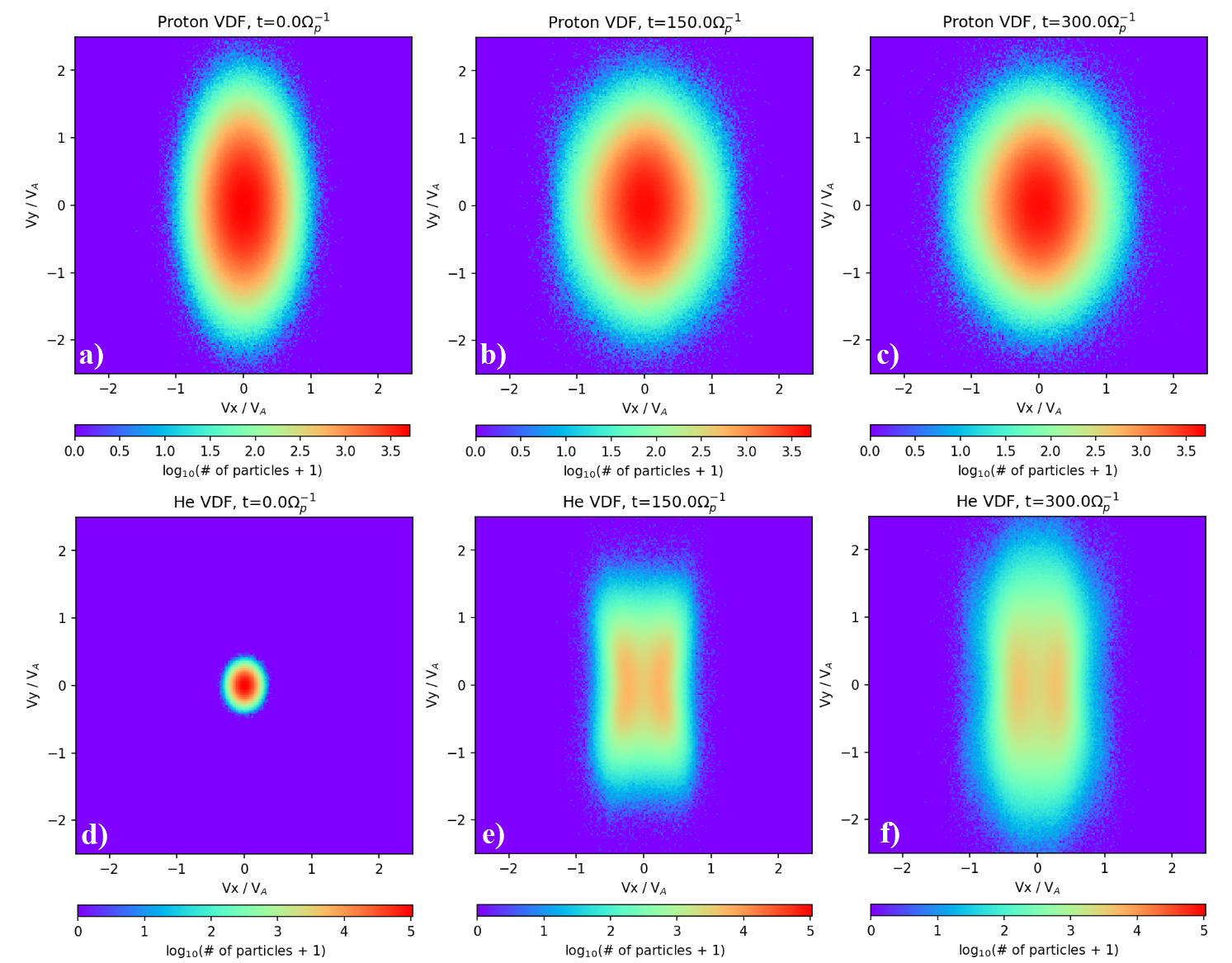}
\end{minipage} \\
\begin{minipage}{1.00\linewidth}
    \centering\includegraphics[width=0.45\linewidth]{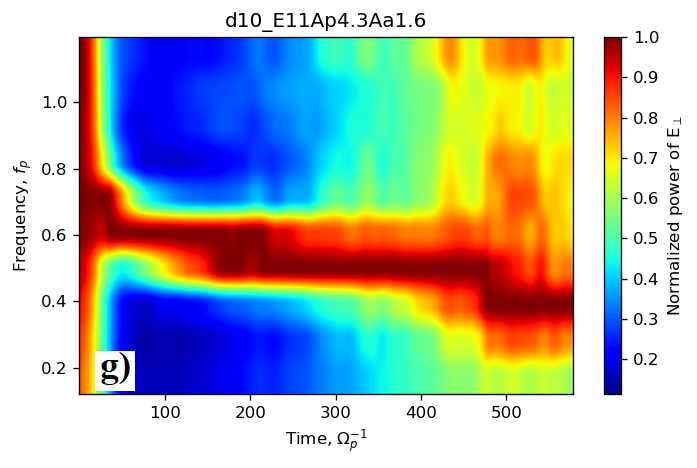}
    \centering\includegraphics[width=0.45\linewidth]{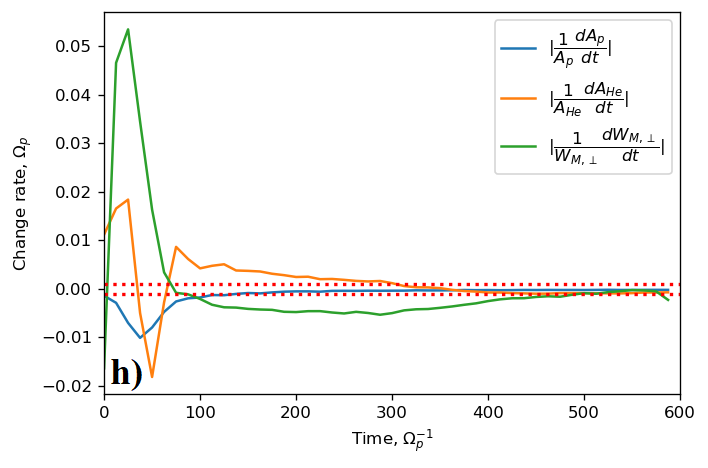}
\end{minipage}
\caption{Examples of proton and $\alpha$ VDFs integrated over $v_{z}$ for simulation run \#4 in Table~\ref{table:hybrid-pic-runs}, at $t=$0\,$\Omega_{p}^{-1}$ (panels a and d), $t=$150\,$\Omega_{p}^{-1}$ (panels b and e), and $t=$300\,$\Omega_{p}^{-1}$ (panels c and f). 
The external magnetic field is directed along the x-axis (it has a non-zero x-component). 
The velocities of particles are normalized to $V_{A}$.
Panel (g) displays the power spectrum of $E_{\perp}$ normalized for every time moment separately along the duration of the run.
Panel (h) displays the change rates of the anisotropies, $A_{p}$ and $A_{He}$, and the perpendicular magnetic energy, $W_{M,\perp}$.
Red dotted lines correspond to the empirical instability threshold $\pm{}0.001\Omega_{p}$ used in this study.
The cases for all three time moments are therefore unstable.}
\label{figure:VDFexamples}
\end{figure}

Hybrid-PIC simulations `track' in time and space the velocities of the $p$ and $\alpha$ particles explicitly, therefore providing a possibility to construct the VDFs and compute the related properties such as the VDF moments along and across the external magnetic field axis. 
Examples of the VDFs for the simulation run \#4 at two time moments are presented in Figure~\ref{figure:VDFexamples}a-f with a high velocity resolution of 0.025$V_{A}$. 
For each simulation snapshot sampled, the following properties were constructed or computed from the VDFs and the corresponding magnetic fields:

\begin{itemize}
    \item Moments of the VDFs computed as $M_{k}=\int{}(v-\bar{v})^{k}f(v)dv/\int{}f(v)dv$ for $k\in{}\{2,3,4\}$, and $M_{1}=\int{}vf(v)dv/\int{}f(v)dv$, for the $p$ and $\alpha$ particles;
    \item Total and perpendicular magnetic energies, $W_{M}=\int{}_{V}\frac{B^2}{8\pi}dV$ and $W_{M,\perp}=\int{}_{V}\frac{B_{\perp}^2}{8\pi}dV$, and related temporal change rates;
    \item Temperature anisotropies of the proton and $\alpha$ particle distributions, $A_{p,\alpha}=(T_{\perp}/T_{\parallel})_{p,\alpha}$, and the related temporal rates of change;
    \item VDFs integrated over $v_{z}$ for the $p$ and $\alpha$ particles with the velocity space resolutions reduced to 0.15$V_{A}$ and 0.10$V_{A}$ respectively.
\end{itemize}

In addition to the properties of the VDFs, we investigate whether the information about the electromagnetic wave spectrum accompanying the instability development could help with the identification of the instabilities. 
The relevant kinetic wave activity produces perturbations in the direction perpendicular to the background magnetic field (i.e., transverse waves), while the propagation direction of these waves can be parallel or oblique. 
Therefore, we first interpolate the perpendicular electric field in the simulations, $E_{\perp}$, in each spatial grid point to the constant temporal grid with the time step of 2.5$\Omega_{p}^{-1}$. 
We note here that the information about the electric and magnetic fields is typically saved in the simulations with a much smaller time step with respect to the particle data, and the time step of 2.5$\Omega_{p}^{-1}$ represents the largest among such time steps of the performed simulations. 
Second, we perform a Fast Fourier Transform (FFT) with the Hanning window at each spatial grid point within the sliding temporal window of 50$\Omega_{p}^{-1}$ centered at every considered time moment of the simulation, and compute the corresponding power spectrum. 
Third, we average the power spectrum computed at each spatial grid point across the domain for every time moment. 
The result represents the wave power in $E_{\perp}$ at 10 frequencies, with the highest frequency slightly larger than the proton gyrofrequency (non-cyclic), $f_{max}\approx{}1.26\,f_{p}$, and the frequency resolution of $\Delta{}f\approx{}0.126\,f_{p}$. 
An example of the dynamic power spectra normalized to unity for each time step for simulation run \#4 is presented in Figure~\ref{figure:VDFexamples}g.
We note here that the wide power spectrum at the beginning of the simulations is a result of the normalization: there is still no distinct $E_{\perp}$ fluctuations developed at the initial moments, and the non-normalized spectrum is mostly dominated by the random perturbations in $E_{\perp}$.

\subsection{Setup of the instability condition}

The goal of this study is to understand how ML could be utilized for the identification of the instability developing in Hybrid-PIC simulations with the ultimate goal of developing the technique for application to PSP, Solar Orbiter, and similar spacecraft data. 
Typically, the stability of VDFs is established by the linear theory from the linearized Vlasov's equations, as the presence of exponentially growing solutions for the electric/magnetic fields for the perturbations at particular frequency ranges. 
However, it becomes harder to define an ``unstable'' or ``stable'' plasma once the simulations reach the non-linear saturation regime: the wave activities are present in the simulations, yet the related VDFs do not evolve as the system reaches a quasi-steady state with no net energy transfer between the waves and particles. 
Therefore, in this work, we adopt an ad-hoc, but physically motivated, definition of an ``unstable'' state of the simulated solar wind plasma, applicable to the nonlinear regime, described below.

We consider the temporal evolution of the temperature anisotropies, $A_{p,\alpha}=T_{\perp,p,\alpha}/T_{||,p,\alpha}$, and perpendicular magnetic energy, $W_{M,\perp}=\int{}_{V}\frac{B_{\perp}^2}{8\pi}dV$. 
The sample of a VDF in the simulation run is considered to be ``unstable'' if the change rate of any of these quantities exceeds a certain threshold. 
For the extensive testing of various machine learning approaches in this work, we use $\left\lvert{}\dfrac{1}{A}\dfrac{dA}{dt}\right\rvert{}<0.001\Omega{}_{p}$ and $\left\lvert{}\dfrac{1}{W_{M,\perp}}\dfrac{dW_{M,\perp}}{dt}\right\rvert{}<0.001\Omega{}_{p}$  as the stability condition. 
The selection of this particular value of the change rate is motivated by the instability threshold growth rate for anisotropy-driven linear stability analysis that is commonly set to $\gamma{}=0.001\Omega{}_{p}$ \citep[see, e.g.,][]{Hellinger2006JGRA..111.1107H,Bale2009PhRvL.103u1101B,Verscharen2016collisionless}.
We note that \citet{Gary1994JGR....99.5903G,Gary1998JGR...10314567G} have investigated a range of instability thresholds with growth rate $\gamma{}$ in the range $\gamma{}\in[10^{-4};10^{-2}]$ for ion cyclotron and resonant firehose instabilities by comparing to the results of nonlinear hybrid simulations.
The instability threshold growth rate curves in the $A_{p}$-$\beta{}_{||}$ plane derived from simulations were consistent with the predictions of the linear stability analysis and did not vary strongly with the change of $\gamma{}$ in the above small growth-rate range.
We also later consider other thresholds, \{$0.01\Omega{}_{p}$, $0.005\Omega{}_{p}$, $0.001\Omega{}_{p}$, $0.0005\Omega{}_{p}$, $0.0001\Omega{}_{p}$\}, but only for the best-performing machine learning approaches. 
To give an idea of the resulting dataset, for the thresholds of instabilities \{$0.01\Omega{}_{p}$, $0.005\Omega{}_{p}$, $0.001\Omega{}_{p}$, $0.0005\Omega{}_{p}$, $0.0001\Omega{}_{p}$\}, the corresponding numbers of the unstable cases are \{62, 110, 418, 681, and 1363\} out of 1596 total VDF samples considered. 
We discuss the separation of the resulting dataset into the train-test subsets, as well as the adopted ML algorithms, in Section~\ref{sec:MLmodel}.
We note that, while one can combine different thresholds for the magnetic energy and anisotropy change rates separately, using the same threshold for both parameters provides a relatively good agreement.
For example, the threshold of $0.001\Omega{}_{p}$ mostly used in this work leads to 418 unstable cases out of 1596 overall. Among them, 381 cases exceed this threshold for the perpendicular magnetic energy change rate, 243 exceed for the anisotropy change rate (for either protons or alpha particles), and 206 exceed the threshold in both considered parameters.
We also note that the consideration of other physical parameters describing the evolution of the VDFs and waves, such as entropy, and the combinations of various thresholds for the instability identification, remain open questions and deserve further investigation.

\section{Machine Learning Pipeline}\label{sec:MLmodel}

The instability identification problem can be viewed as a binary classification problem once the definition of instability is specified. 
The corresponding inputs into the ML algorithm could be either the statistical moments of the VDFs augmented with other parameters of the simulation runs (such as the relative abundances of ions) or the VDFs directly as two-dimensional arrays also augmented with relative abundances. 
This gives us an opportunity to consider the ML binary classification algorithms that utilize the features of the VDFs (the statistical moments in this study) and compare them with the image-centered deep learning approaches.

\subsection{Feature-Based Classification}

For the feature-based classification problem, we consider four widely used ML classification methods, all implemented within the {\tt scikit-learn} Python library package \citep{scikit-learn}: 
\begin{itemize}
    \item k-Nearest Neighbor (kNN) classifier.
    The classifier assigns a new case to be stable or unstable depending on its k closest neighbors-cases found based on a chosen distance metric.
    For a more expanded description of the kNN, please see, e.g., \citet{Nishizuka2017ApJ...835..156N};
    \item Support Vector Machine (SVM) classifier with the Radial Basis Function (RBF) kernel \citep{Cortes1995support}.
    The classifier maps input cases into a higher-dimensional space to find an optimal hyperplane that maximizes the margin between stable and unstable cases.
    A more detailed description of RBF SVM can be found in, e.g., \citet{Sadykov2017ApJ...849..148S};
    \item Multi-Layer Perceptron (MLP) classifier with one hidden layer \citep{Cybenko1989MCSS....2..303C}.
    MLP is a feedforward neural network that can approximate any nonlinear relationship given a sufficient number of neurons in the hidden layer.
    For more description of MLPs, please see, e.g., \citet{Goodwin2024ApJ...964..163G};
    \item Random Forest (RF) classifier \citep{Breiman2001MachL..45....5B}.
    RF is an ensemble learning method that constructs multiple decision trees during training, each utilizing a different subset of features as input, and outputs the majority vote of their predictions.
    More information about the Random Forest and individual decision trees can be found in, e.g., \citet{Breiman2001MachL..45....5B,Goodwin2024ApJ...964..163G}. 
\end{itemize}

While there exists a much wider variety of possible methods to employ, the selected methods represent a good pool of approaches within the scope of our project, from the approaches relying on the behavior of samples in the neighborhood with similar properties (kNN, SVM RBF) to ones capable of capturing complex non-linear dependencies \citep[MLP][]{Cybenko1989MCSS....2..303C}, or representing the ensemble decision of the individually-weak learners (RF). 
During the processing of the simulation runs, we computed four statistical moments for the directions along and across the external magnetic field for both $p$ and $\alpha$ particles, which resulted in 16 different parameters/features to be utilized for ML. 
Together with the temperature anisotropies and relative particle population abundances for $p$ and $\alpha$, this gives a total of 20 features as an entry for the classifier. 
In case the $E_{\perp}$ power spectrum is considered, we also add the normalized powers at each of the 10 frequencies, raising the total number of features to 30. 
For every feature, we perform the standard scaling normalization, since most of the considered ML algorithms (kNN, SVM, MLP) are sensitive to the normalization scales.

For every considered ML algorithm, several hyperparameters are required to be optimized. 
We perform here the 10-fold cross-validation to determine which are the best parameters. 
The optimization was done with respect to the accuracy score described further in the text. 
The corresponding hyperparameters considered (with the most optimal indicated by a single asterisk for the case with no dynamic power spectrum considered, and by a double asterisk for the case with the dynamic power spectrum involved in training) are:
\begin{itemize}
    \item kNN: \verb|`n_neighbors'|: \{2$^{**}$, 5$^{*}$, 10, 25\}, \verb|p|: \{1$^{*,**}$, 2\};
    \item SVM: \verb|`C'|: \{0.001, 0.01, 0.1, 1, 10, 100, 1000, 10000$^{*,**}$\}, \verb|`gamma'|: \{0.0001, 0.001$^{**}$, 0.01, 0.1, 1, 'scale'$^{*}$\}, \verb|`class_weight'|: \{None$^{*,**}$, 'balanced'\};
    \item MLP: \verb|`hidden_layer_sizes'|: \{10, 50, 200$^{**}$, 500$^{*}$\}, \verb|`max_iter'|: \{100, 250$^{**}$, 1000$^{*}$, 2500, 10000\};
    \item RF: \verb|'n_estimators'|: \{10, 50, 100$^{**}$, 200$^{*}$\}, \verb|`max_depth'|: \{None$^{*}$, 2, 5, 10, 25$^{**}$\}, \verb|`class_weight'|: \{None$^{*,**}$, `balanced'\}
\end{itemize}

\subsection{VDF-Based Classification}

An alternative approach is to input the VDFs directly into the classification model. 
For this purpose, we employ the Convolutional Neural Networks \citep[CNNs,][]{krizhevsky2012imagenet}.
We note here that the CNN approach has previously been applied on the synthetic VDFs for retrieving the properties of the proton core and beam components, and $\alpha$-particle parameters \citep{Vech2021A&A...650A.198V}.
The VDFs of the protons and $\alpha$ particles are reduced in resolution in velocity space (we experiment with the resolutions $0.15V_{A}$ and $0.10V_{A}$), normalized to one in the phase density, and stacked together as two input channels.
To introduce relative populations of the $p$ and $\alpha$ particles, as well as the $E_{\perp}$ power spectrum (if considered), we concatenate them with the flattened output of the convolutional layers. 
As a result, two input channels are propagating to the CNN, and the outputs of the convolutional layers are flattened and concatenated with the population and power spectrum information before propagating into the fully connected part of the network. 
While we do not perform an exhaustive architecture search in this study, we test the following possibilities for the convolutional and fully-connected parts of the network:
\begin{itemize}
    \item The convolutional part of the network can have 1, 2, or 3 convolutional layers. 
    Each convolutional operation has a kernel size of 3 with no stride and doubles the number of channels. 
    Each layer is followed then by the max pooling operation with a kernel size of 2 and a stride of 2, and by the Rectified Linear Unit (ReLU) activation function to introduce non-linearity into the network. 
    The outputs of the last layer are then reshaped into a one-dimensional vector.
    \item The fully-connected part of the network can have 1 or 2 hidden layers, of the number of neurons of 10 or (50,10), correspondingly. 
    Each linear layer is followed by the ReLU activation function, except the output one, which has the Sigmoid activation function (to confine the output between 0 and 1). 
    The network has two output channels that represent the predicted likelihoods of the case being stable or unstable, respectively; the case is categorized based on the higher of the two likelihoods.
\end{itemize}

The networks were implemented using PyTorch \citep{Paszke2019arXiv191201703P}. 
For the network training, the binary cross-entropy loss function was used:
\begin{gather}
\label{eq:crossentropy}
    \mathcal{L} = -\dfrac{1}{N}\sum_{i=1}^{N}y_{i}\cdot{}log(p_{i}) + (1-y_{i})log(1-p_{i}).
\end{gather}

Here $N$ is the total number of samples in the data batch (kept fixed and equal to 128), $y_{i}$ is the label of the $i$-th input (1 or 0), and $p_{i}$ is the predicted probability of the $i$-th input to have a label 1. 
The ``adam'' optimizer \citep{Kingma2015} with a learning rate of 0.001 was used to progress the training, and the weight decay of 0.001 to prevent overfitting. 
The training of the network continued for the different number of epochs (250, 500, 750, 1000, 1250, 1500, 1750) which served as a hyperparameter of the network as well, and then the results corresponding to the performance of the network on the evaluation dataset after all the training steps have been recorded.

\begin{figure}[htbp]
\centerline{\includegraphics[width=0.9\linewidth]{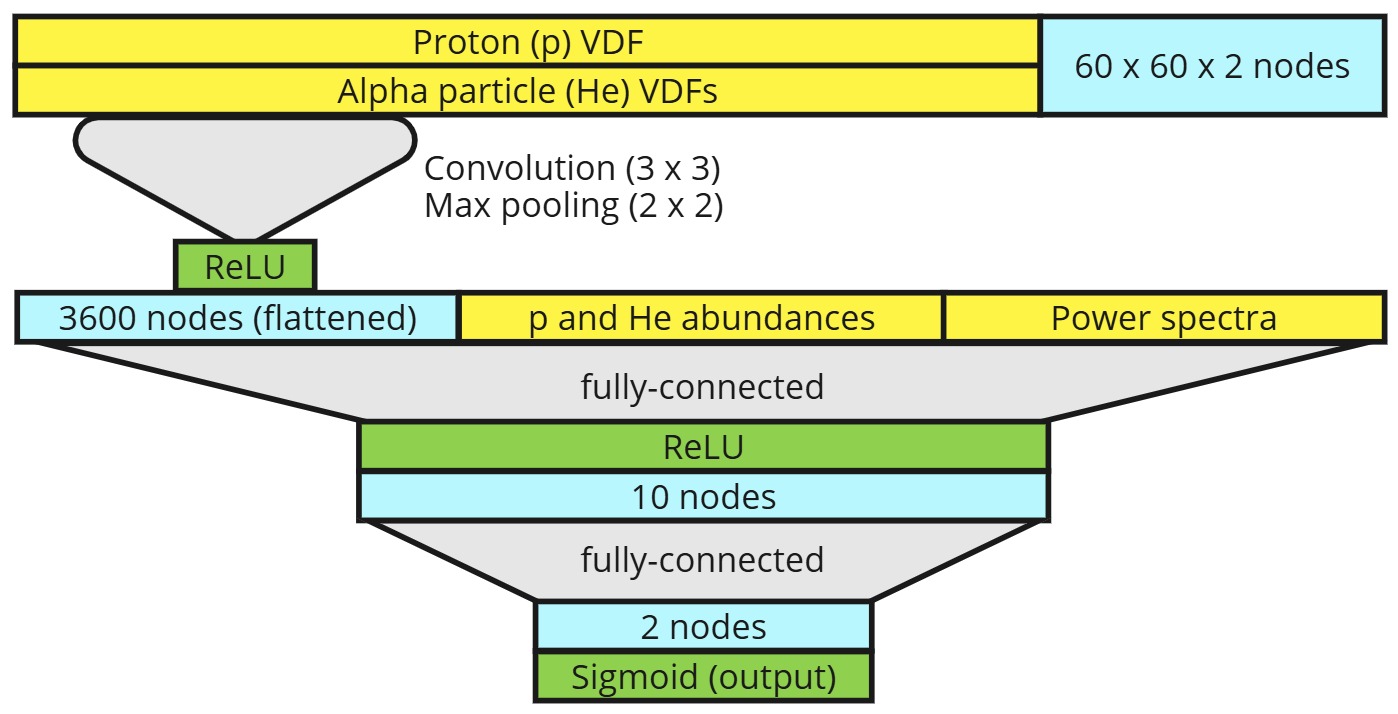}}
\caption{Architecture of the best-performing Convolutional Neural Network (CNN) working with the VDFs of 0.10$V_{A}$ resolution with power spectrum information included.}
\label{figure:cnn_architect}
\end{figure}

As described above, there are 6 different network configurations possible (3 convolutional options $\times$ 2 fully-connected options) for each resolution of the input VDF, $0.15V_{A}$ and $0.10V_{A}$, along with 7 choices for the total number of the training epochs. 
Figure~\ref{figure:cnn_architect} illustrates the architecture that was found to work the best for the VDFs with the 0.10$V_{A}$ resolution for the case when the power spectrum information of the ion-scale waves was utilized. 
As one can see from Figure~\ref{figure:cnn_architect}, the architecture has one convolutional layer and two hidden layers in its fully-connected part. 
The type of architecture (adjusted to the smaller size of the input) was found to be optimal for the 0.15$V_{A}$ velocity space resolution as well. 
The most optimal CNN architectures are also summarized in Table~\ref{table:CNN_arch}.

\begin{table*}[htbp]
\caption{Optimal architectures of the Convolutional Neural Network (CNN).}
\begin{center}
\begin{tabular}{|c|cccc|}
\hline
\textbf{VDF resolution} &   0.10$V_{A}$ &   0.10$V_{A}$ &   0.15$V_{A}$ &   0.15$V_{A}$ \\
\hline
\textbf{$E_{\perp}$ Power spectrum} &   No  &   Yes &   No  &   Yes \\
\textbf{Epochs} &   1750    &   1750    &   1000    &   1750    \\
\textbf{\# Convolutional Layers}    &   1   &   1   &   1   &   1   \\
\textbf{\# Fully-connected layers}  &   1   &   2   &   1   &   2   \\
\hline
\end{tabular}
\label{table:CNN_arch}
\end{center}
\end{table*}

\subsection{Assessment of Binary Classification Results}

The outcome of every binary classification problem results represents four numbers: the number of True Positive predictions (TP), True Negative predictions (TN), False Positive predictions (FP), and False Negative predictions (FN). 
In this work, the TP means that the unstable distribution/case is predicted correctly as unstable, TN~--- stable is predicted correctly as stable, FP~--- stable is predicted incorrectly as unstable, and FN~--- unstable is predicted incorrectly as stable.
For convenience, we also summarize the definitions in Table~\ref{table:confusionmatrix}.
From these four numbers, the assessment metrics can be constructed. We will consider two of them in this work, the accuracy and the True Skill Score (TSS) defined as:
\begin{gather}
    \textrm{Accuracy} = \dfrac{TP+TN}{TP+FP+FN+TN} \\
    \textrm{TSS} = \dfrac{TP}{TP+FN} - \dfrac{FP}{FP+TN}
\end{gather}

\begin{table*}[htbp]
\caption{The results of the classification problem (also known as confusion matrix) and their meaning within this work.}
\begin{center}
\begin{tabular}{|c|c|c|}
\hline
\textbf{TP}  &   \# of True Positive predictions   &   unstable cases predicted correctly as unstable   \\
\hline
\textbf{TN}  &   \# of True Negative predictions   &   stable cases predicted correctly as stable   \\
\hline
\textbf{FP}  &   \# of False Positive predictions   &   stable cases predicted incorrectly as unstable   \\
\hline
\textbf{FN}  &  \# of  False Negative predictions   &   unstable cases predicted incorrectly as stable   \\
\hline
\end{tabular}
\label{table:confusionmatrix}
\end{center}
\end{table*}

The accuracy measures the fraction of the correct predictions over the total number of predictions. 
However, it is known to misrepresent the prediction effort in the cases of highly imbalanced datasets \citep[see, for example,][]{OKeefe2024-RHclassifier}. 
Therefore, we consider another metric, TSS, which is insensitive to the class imbalance of the dataset in the sense that the duplication of the positive or negative samples does not change the score. 
The TSS metrics are widely used in space weather forecasting tasks \citep{Goodwin2024ApJ...964..163G,Sadykov2017ApJ...849..148S}. 
The scores are averaged over the train-validation pairs, and the corresponding standard deviations are computed.

\section{Results}\label{sec:results}

\begin{figure*}[htbp]
\centerline{\includegraphics[width=0.5\linewidth]{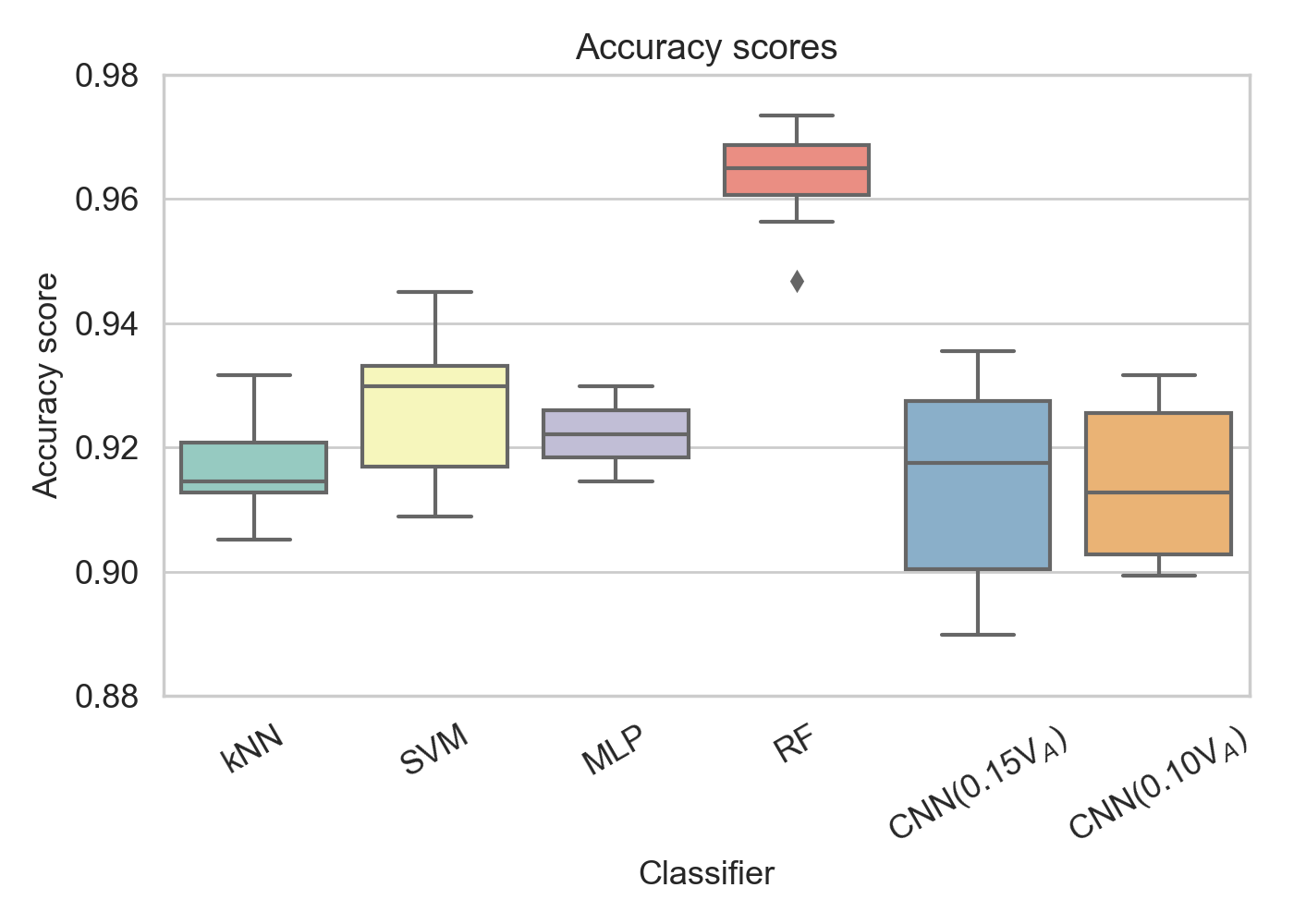}
\includegraphics[width=0.5\linewidth]{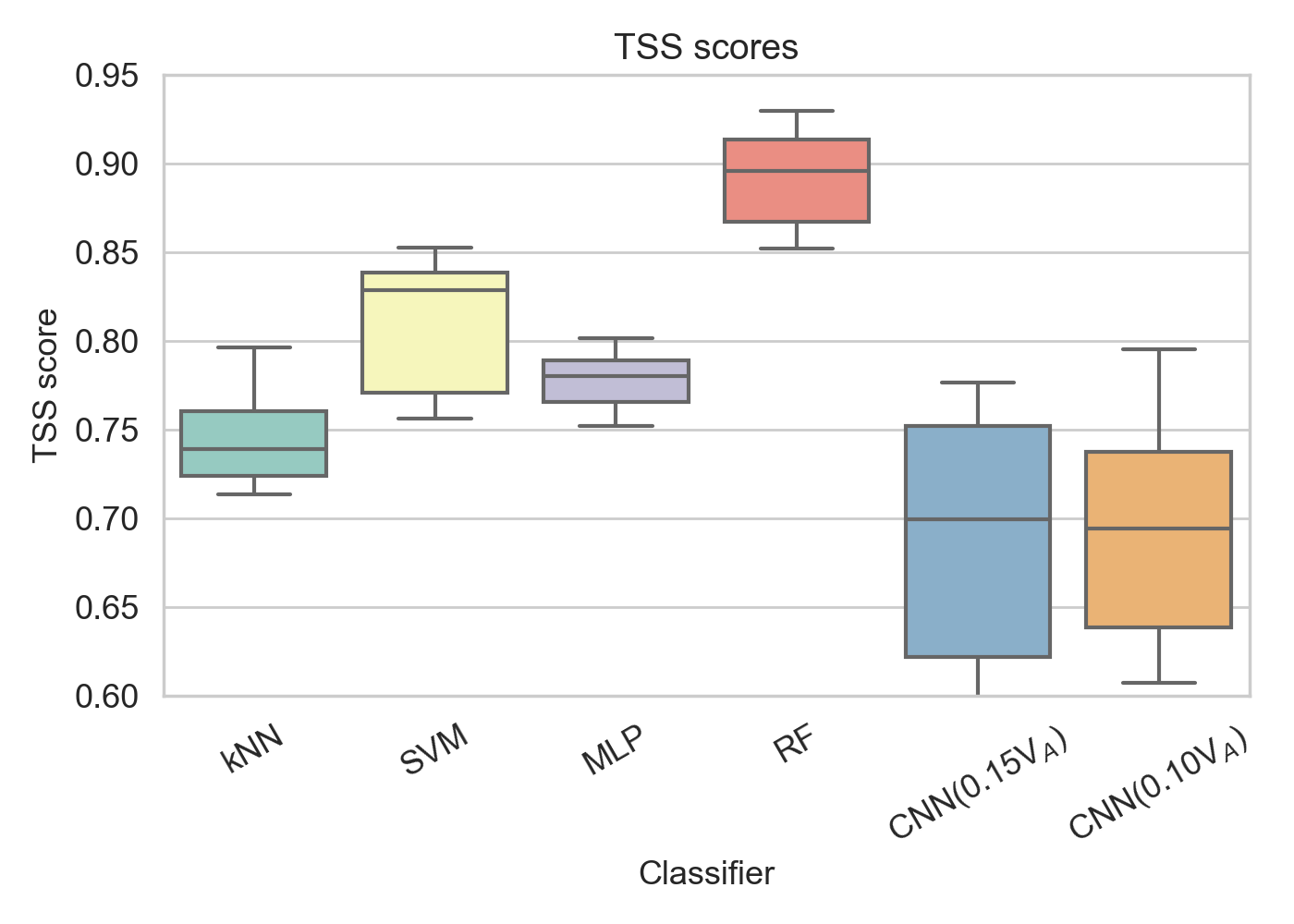}}
\caption{Box-and-whiskers plots summarizing the accuracy scores (left) and the TSS scores (right) of binary classification models considered in this work, over 10 randomized train-test data splits. 
Each colored rectangle spans through the second and third quartiles of the scores, the horizontal bar marks the median. 
The whiskers indicate the locations of the last individual results within the interquartile range from the boxes. 
The rhombus points mark outliers outside the interquartile range from the boxes.}
\label{figure:scores}
\end{figure*}

The summary of the binary classification results for the stability threshold of $<0.001\Omega_{p}$ (i.e., when the perpendicular magnetic energy and $p$ and $\alpha$ temperature anisotropy relative change rates are slower than this specified rate) for all classification models without the use of the $E_{\perp}$ power spectrum information are presented in Table~\ref{table:results_thres0001} in terms of the mean scores and standard deviations across 10 randomized train-validation splits. 
The train-validation splits were kept the same for the considered models, ensuring identical data conditions for evaluation. 
For comparison, the table also indicates the performance of the trivial forecasts, random-chance and always-negative, for the entire dataset.
The random-chance forecast is a forecast when each case has a 50\% probability of being classified as stable or unstable. The always-negative forecast corresponds to the forecast when every case is predicted as stable.
In addition, the box-and-whisker plots summarizing 10 experiments are presented in Figure~\ref{figure:scores} for the accuracy and TSS scores of the models.

\begin{table*}[htbp]
\caption{Results of the instability identification for the threshold of $0.001\Omega{}_{p}$. 
The last two lines in the table summarize the random-chance forecast and always-negative forecast performances.}
\begin{center}
\begin{tabular}{|c|cccccc|}
\hline
\textbf{ML Model} &   \textbf{TP}   &   \textbf{TN} &   \textbf{FP}  &   \textbf{FN}  &   \textbf{Accuracy} &   \textbf{TSS} \\
\hline
kNN &   105$\pm$5   &   378$\pm$3   &   15$\pm$3    &   29$\pm$3    &   0.92$\pm$0.01   &   0.74$\pm$0.02   \\
SVM RBF &   116$\pm$7   &   372$\pm$5   &   20$\pm$3    &   19$\pm$5    &   0.93$\pm$0.01   &   0.81$\pm$0.04   \\
MLP &   110$\pm$5   &   376$\pm$5   &   17$\pm$4    &   24$\pm$3    &   0.92$\pm$0.01   &   0.78$\pm$0.02   \\
Random Forest &   122$\pm$6   &   386$\pm$5   &   7$\pm$3    &   12$\pm$4    &   0.96$\pm$0.01   &   0.89$\pm$0.03   \\
CNN (0.15$V_{A}$) &   94$\pm$10   &   387$\pm$6   &    6$\pm$4    &   40$\pm$11    &   0.91$\pm$0.02   &   0.69$\pm$0.07   \\
CNN (0.10$V_{A}$) &   95$\pm$10   &   387$\pm$7   &   6$\pm$4    &   39$\pm$9    &   0.91$\pm$0.01   &   0.69$\pm$0.06 \\
\hline
Random Chance   &   209 &   589    &    589   &    209   &  0.50 &   0.0    \\
Negative    &    0   &    1178   &   0    &   418    &   0.74    &    0.0   \\
\hline
\end{tabular}
\label{table:results_thres0001}
\end{center}
\end{table*}

There are several takeaway points from the Table~\ref{table:results_thres0001}. 
First, all considered models demonstrate a high accuracy of 0.91 or greater. 
This value is not only significantly larger with respect to the random chance forecast (accuracy of 0.50) but also larger than for the always-negative prediction (the case when the model always predicts that the case as stable), which would correspond to an accuracy of $\sim$0.74 for the $0.001\Omega_{p}$ threshold. 
Therefore, we are observing that all models are performing better with respect to these baselines. 
All models also demonstrate TSS scores of 0.69 or higher (note that the TSS score is zero by its construction for both the random-chance and always-negative forecasts).

The second important takeaway point from Table~\ref{table:results_thres0001} and Figure~\ref{figure:scores} is that the Random Forest-based model outperforms all other classifiers considered in this work. 
There is a significant increase in accuracy with respect to the second-performing model, which is SVM RBF, from 0.93$\pm$0.01 to 0.96$\pm$0.01. 
The same occurs to the TSS metrics: Random Forest has a TSS score of 0.89$\pm$0.03 compared to a score of  0.81$\pm$0.04 for the second-best SVM RBF model. 
By looking at these scores, we can conclude that the rest of the feature-based models (kNN, SMV RBF, and MLP) intersect in terms of their performance: the corresponding scores overlap within their standard deviations across the train-validation partitions. 
The possible reasons for such behavior of Random Forest, and the corresponding implications to instability identification, are discussed in Section~\ref{sec:discussion}.

We also note here that while the performance of the CNN models was comparable to others in terms of accuracy, the corresponding TSS values (TSS$=$0.69$\pm$0.07 and TSS$=$0.69$\pm$0.06 for the VDFs fed with 0.15$V_{A}$ and 0.10$V_{A}$ velocity space resolutions, correspondingly) were significantly lower than for the weakest performing feature-based model, kNN (TSS$=$0.74$\pm$0.02). 
Notably, the CNN models resulted in a lower number of TP predictions and a higher number of FN predictions. 
Both Table~\ref{table:results_thres0001} and Figure~\ref{figure:scores} indicate that the CNN models experience a larger spread of their performance metrics across the train-validation pairs, indicating their higher sensitivity to the particular training dataset.

\begin{table*}[htbp]
\caption{Same as Table~\ref{table:results_thres0001} but with the inclusion of the dynamic power spectrum information.}
\begin{center}
\begin{tabular}{|c|cccccc|}
\hline
\textbf{ML Model} &   \textbf{TP}   &   \textbf{TN} &   \textbf{FP}  &   \textbf{FN}  &   \textbf{Accuracy} &   \textbf{TSS} \\
\hline
kNN & 108$\pm$7 & 387$\pm$5 & 6$\pm$2 & 27$\pm$5 & 0.94$\pm$0.01 & 0.79$\pm$0.03 \\
SVM RBF & 117$\pm$6 & 379$\pm$6 & 13$\pm$4 & 18$\pm$4 & 0.94$\pm$0.01 & 0.83$\pm$0.03 \\
MLP & 115$\pm$6 & 379$\pm$5 & 13$\pm$3 & 19$\pm$5 & 0.94$\pm$0.01 & 0.82$\pm$0.03 \\
Random Forest & 121$\pm$5 & 386$\pm$5 & 7$\pm$3 & 13$\pm$4 & 0.96$\pm$0.01 & 0.88$\pm$0.03 \\
CNN (0.15$V_{A}$) & 102$\pm$9 & 384$\pm$7 & 9$\pm$6 & 33$\pm$8 & 0.92$\pm$0.02 & 0.73$\pm$0.06 \\
CNN (0.10$V_{A}$) & 99$\pm$14 & 388$\pm$7 & 5$\pm$4 & 36$\pm$10 & 0.92$\pm$0.02 & 0.72$\pm$0.08 \\
\hline
Random Chance   &   209 &   589    &    589   &    209   &  0.50 &   0.0    \\
Negative    &    0   &    1178   &   0    &   418    &   0.74    &    0.0   \\
\hline
\end{tabular}
\label{table:results_thres0001_ps}
\end{center}
\end{table*}

\begin{figure}[htbp]
\centerline{\includegraphics[width=0.5\linewidth]{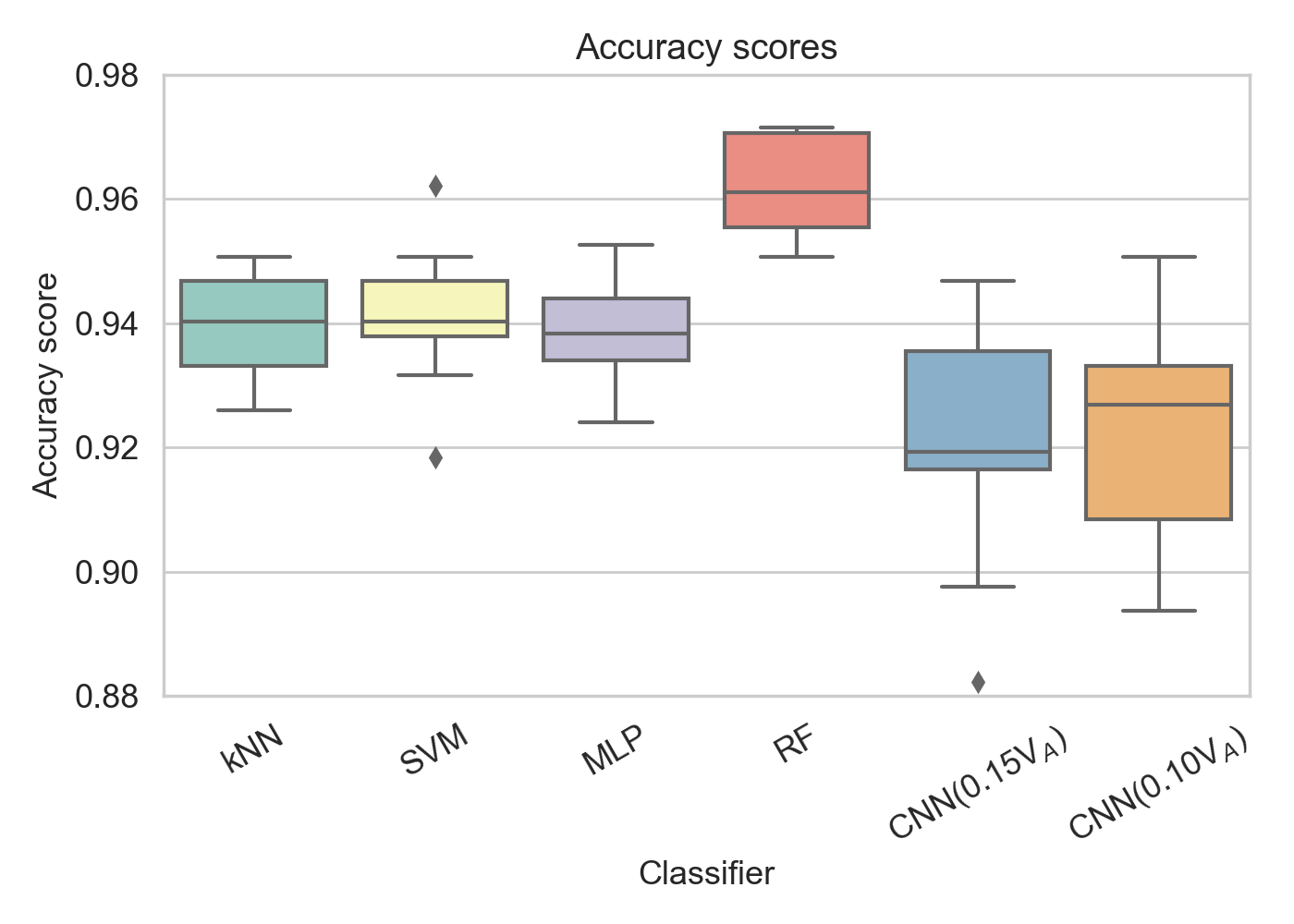}
\includegraphics[width=0.5\linewidth]{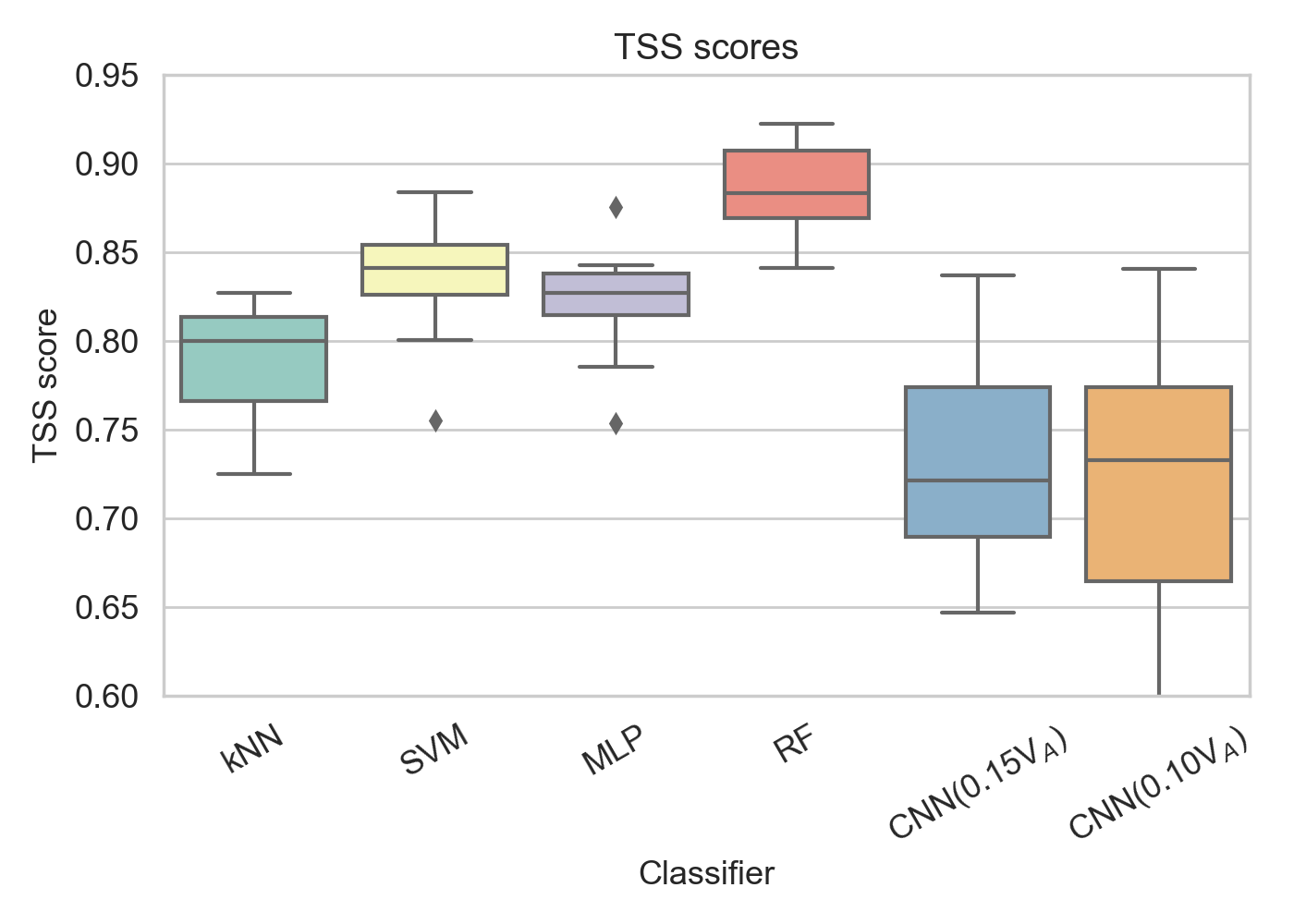}}
\caption{Same as Figure~\ref{figure:scores} but with the inclusion of the dynamic power spectrum information.}
\label{figure:scores_ps}
\end{figure}

Table~\ref{table:results_thres0001_ps} and Figure~\ref{figure:scores_ps} illustrate the performance of various ML/DL models in the situation when the $E_{\perp}$ power spectrum information was fed into the classifiers. 
The performance was assessed across exactly the same train-validation partitions as for Table~\ref{table:results_thres0001}. 
One can notice that the performance of all models has increased with respect to the case where $E_{\perp}$ power spectrum information was not used, except for the Random Forest (stayed the same within the cross-validation uncertainty). 
For example, for the MLP model, the accuracy increased from 0.92$\pm$0.01 to 0.94$\pm$0.01, and the TSS score changed from 0.78$\pm$0.02 to 0.82$\pm$0.03. 
The CNN models also demonstrated a slight improvement. 
Overall, we can conclude that the performance scores of the models were either unchanged (in the case of Random Forest) or increased (in the case of all other models) following the incorporation of the $E_{\perp}$ power spectrum.

\begin{table*}[htbp]
\caption{Performance of the Random Forest classifier on the varying instability threshold (hyperparameter optimization done with respect to TSS score). 
The last column represents the class imbalance in the dataset (number of stable cases VS number of unstable cases).}
\begin{center}
\begin{tabular}{|c|ccccccc|}
\hline
\textbf{Threshold} &   \textbf{TP}   &   \textbf{TN} &   \textbf{FP}  &   \textbf{FN}  &   \textbf{Accuracy} &   \textbf{TSS} & \textbf{Class-imb.} \\
\hline
$>0.0001\Omega_{p}$ &   417$\pm$9   &   75$\pm$7   &   4$\pm$1    &   30$\pm$7    &   0.93$\pm$0.01   &   0.88$\pm$0.02  & 233:1363 \\
$>0.0005\Omega_{p}$ &   207$\pm$6   &   294$\pm$6   &   11$\pm$5    &   16$\pm$5    &   0.95$\pm$0.01   &   0.90$\pm$0.02  & 915:681 \\
$>0.001\Omega_{p}$ &   122$\pm$6   &   386$\pm$5   &   7$\pm$3    &   12$\pm$4    &   0.96$\pm$0.01   &   0.89$\pm$0.03  & 1178:418 \\
$>0.005\Omega_{p}$ &   30$\pm$5   &   475$\pm$5   &    17$\pm$3    &   6$\pm$2    &   0.96$\pm$0.01   &   0.81$\pm$0.07  & 1486:110 \\
$>0.01\Omega_{p}$ &   16$\pm$2   &   496$\pm$6   &    11$\pm$6    &   5$\pm$3    &   0.97$\pm$0.01   &   0.75$\pm$0.11  & 1534:62 \\
\hline
\end{tabular}
\label{table:results_thresvar}
\end{center}
\end{table*}

\begin{table*}[htbp]
\caption{Same as in Table~\ref{table:results_thresvar} but with the inclusion of dynamic power spectrum information.}
\begin{center}
\begin{tabular}{|c|ccccccc|}
\hline
\textbf{Threshold} &   \textbf{TP}   &   \textbf{TN} &   \textbf{FP}  &   \textbf{FN}  &   \textbf{Accuracy} &   \textbf{TSS} & \textbf{Class-imb.} \\
\hline
$>0.0001\Omega_{p}$ &   426$\pm$10   &   74$\pm$6   &   5$\pm$3    &   21$\pm$7    &   0.95$\pm$0.01   &   0.89$\pm$0.03  & 233:1363 \\
$>0.0005\Omega_{p}$ &   207$\pm$5   &   294$\pm$7   &   11$\pm$4    &   15$\pm$4    &   0.95$\pm$0.01   &   0.90$\pm$0.02  & 915:681 \\
$>0.001\Omega_{p}$ &   121$\pm$6   &   386$\pm$5   &   7$\pm$3    &   13$\pm$3    &   0.96$\pm$0.01   &   0.88$\pm$0.03  & 1178:418 \\
$>0.005\Omega_{p}$ &   29$\pm$4   &   477$\pm$5   &    15$\pm$4    &   6$\pm$4    &   0.96$\pm$0.01   &   0.79$\pm$0.09  & 1486:110 \\
$>0.01\Omega_{p}$ &   15$\pm$2   &   498$\pm$6   &    9$\pm$6    &   5$\pm$3    &   0.97$\pm$0.01   &   0.73$\pm$0.11  & 1534:62 \\
\hline
\end{tabular}
\label{table:results_thresvar_ps}
\end{center}
\end{table*}

Because the Random Forest model behaved best among the considered methods, we utilized only this model for further investigation of varying the instability thresholds. 
Table~\ref{table:results_thresvar} summarizes the outcomes of using Random Forest, with the optimized hyperparameter configuration for every threshold case with respect to the TSS score. 
Table~\ref{table:results_thresvar_ps} illustrates the same but with the inclusion of $E_{\perp}$ power spectrum information. 
The decision to optimize with respect to TSS was made because of the strong variation of class imbalance with the threshold (see the last column of the Table~\ref{table:results_thresvar}) and non-sensitivity of TSS to class imbalance \citep[in the sense that the duplication of the positive or negative samples in the evaluation dataset will keep the TSS score the same,][]{Bobra2015ApJ...798..135B}. 
One can see that the instability threshold of $>0.0005\Omega_{p}$ leads to the highest $TSS=0.90\pm0.02$ score among the considered thresholds both without and with power spectrum information used, and the threshold of $>0.001\Omega_{p}$ which was adopted for a detailed study has the $TSS=0.89\pm0.03$ and $TSS=0.88\pm0.03$, correspondingly, the second-third largest among the observed thresholds. 
The TSS scores drop significantly with increasing the instability threshold. 
Therefore, we can conclude that the thresholds of $>0.0005\Omega_{p}$ and $>0.001\Omega_{p}$ are close to optimal for the utilized dataset in terms of the maximization of the TSS score. 
We can also see that the inclusion of the $E_{\perp}$ power spectrum generally does not improve the results of the instability detection for the case if Random Forest is used.

As evident in Table~\ref{table:hybrid-pic-runs}, the parameter space for the initialization of the hybrid-PIC simulations is large, which causes several challenges that need to be addressed. 
First, as the hybrid-PIC simulations are computationally costly to perform, we have only 34 simulation runs mostly leveraged from the previous studies. While most of them are initialized with conditions guided by observations of the young solar wind \citep[like in][]{Ofman2022ApJ...926..185O,Ofman2025PIC,Yogesh2025PIC}, this list of simulations is far from being exhaustive. 
Therefore, it is important to test the instability detection model on conditions on which the model has not been trained. 
Second, the samples from the same simulation run might potentially experience the temporal coherence problem \citep{Ahmadzadeh2021ApJS..254...23A}, when the samples that are very close to each other may appear both in the train and validation sets. 
This is not necessarily a negative scenario: one wants the model to perform well on the samples that are close to the samples presented in the training dataset. 
However, one also wants to prevent the model from `memorizing' each particular sample or simulation run and not generalizing to the patterns in the data instead.

\begin{figure*}[htbp]
\centerline{\includegraphics[width=\linewidth]{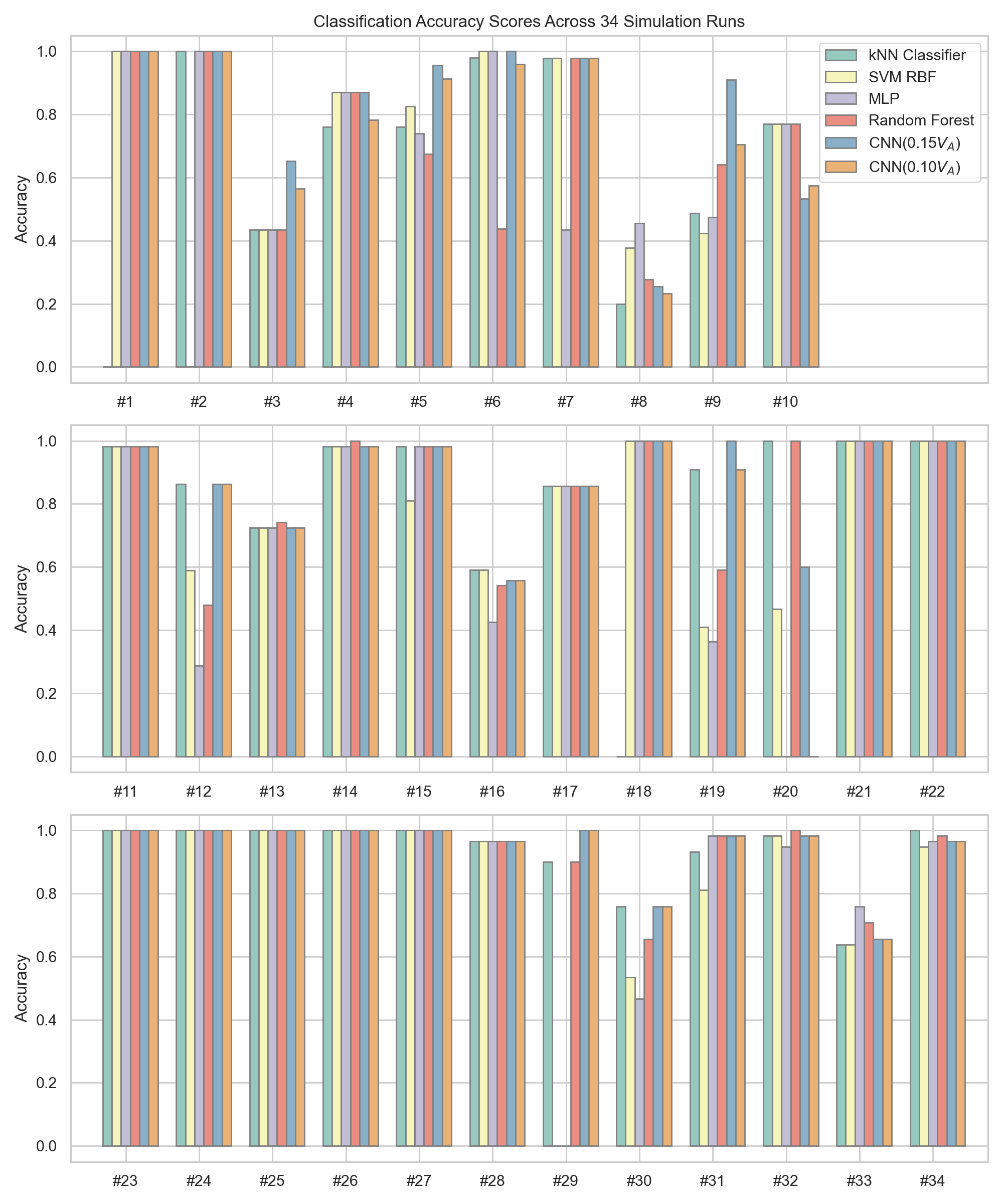}}
\caption{Accuracy of the considered classifiers on the `unseen' hybrid-PIC runs (visualized for each hybrid-PIC run on which the classifiers validate and do not train) for the instability threshold of $0.001\Omega{}_{p}$.
The absence of the bar for a certain ML model and Hybrid-PIC run indicates that its accuracy was zero (i.e., all predictions were incorrect).}
\label{figure:acc_allmodels}
\end{figure*}

\begin{figure*}[htbp]
\centerline{\includegraphics[width=\linewidth]{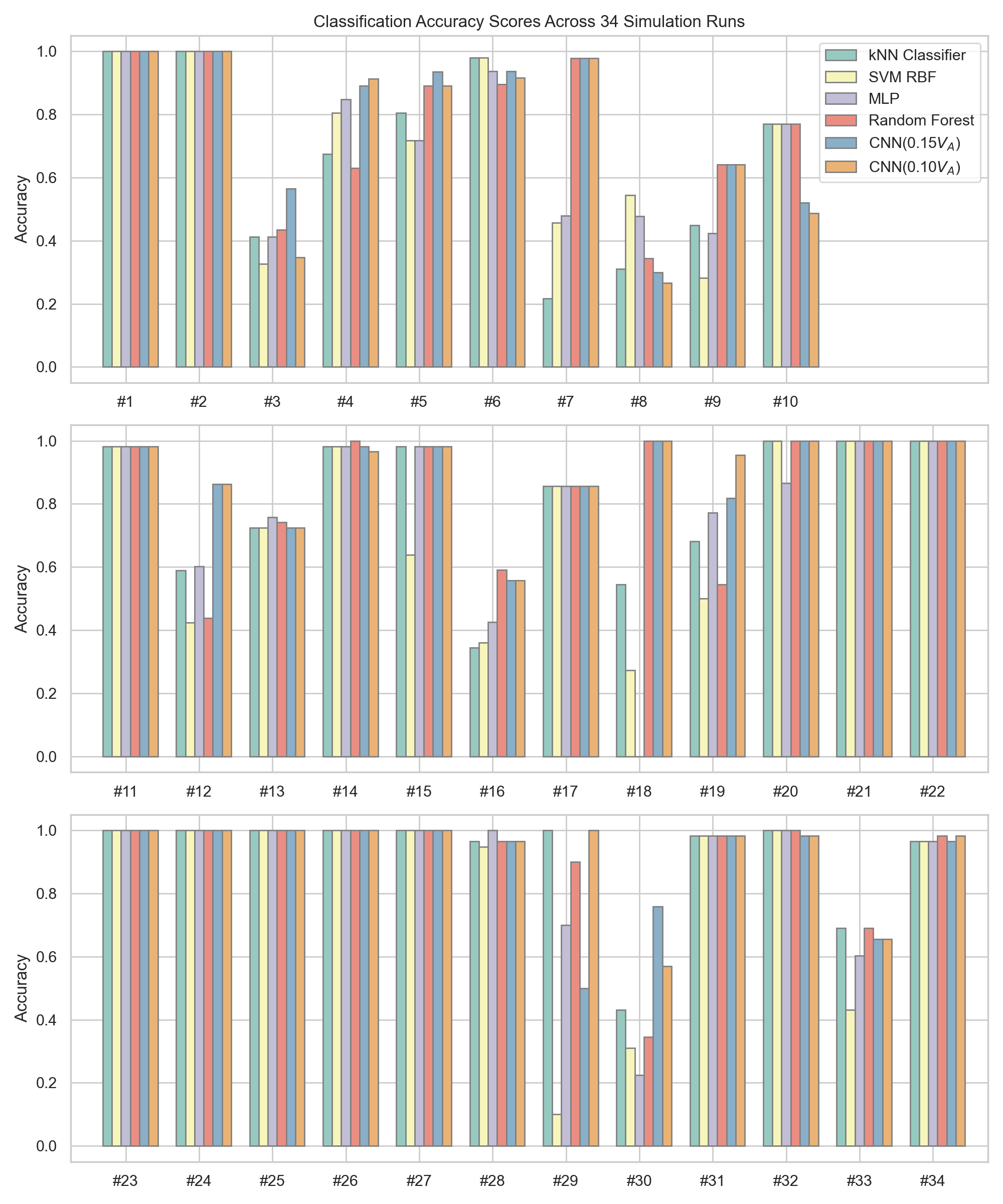}}
\caption{Same as in Figure~\ref{figure:acc_allmodels} but with the inclusion of power spectrum.}
\label{figure:acc_allmodels_ps}
\end{figure*}

Figure~\ref{figure:acc_allmodels} illustrates the experiment when the samples from one of the hybrid-PIC simulation runs were isolated and placed into the validation dataset, and all samples in other simulation runs were used for training the ML model. 
Figure~\ref{figure:acc_allmodels_ps} illustrates the same test but when the information about the $E_{\perp}$ was included in the training process. 
This test illustrates what will happen if the instability detection models see previously `unseen' distributions, and is performed for the instability threshold of $0.001\Omega_{p}$.
For training the models, we are using hyperparameters or configurations that were found to be most optimal for the randomized train-validation split. 
As one can notice, sometimes the detection models perform well irrelevantly of the model type (for example, for the cases \#1. \#2, \#11, \#14, etc.), sometimes only the particular subset of models works well (simulation runs \#7, \#18, \#20, etc). 
However, sometimes the accuracy is relatively low irrespective of a particular model (runs \#3, \#8, \#16, \#30, etc.). 
The potential use of this behavior of the models is discussed in Section~\ref{sec:discussion}. 
One can also notice that the performance of the ML models on unseen runs typically does not depend on whether or not the $E_{\perp}$ power spectrum is used for predictions, remaining more or less consistent, with some individual simulation runs demonstrating consistent improvement in scores among the ML models (such as run \#20), as well as slight decreases in scores (such as runs \#3 and \#30).

One can also compute the average accuracy scores across the models without and with the use of $E_{\perp}$ power spectrum, correspondingly:
\begin{itemize}
    \item kNN: Accuracy$=0.80\pm{}0.27$, $0.80\pm{}0.24$;
    \item SVM RBF: Accuracy$=0.76\pm{}0.28$, $0.74\pm{}0.28$;
    \item MLP: Accuracy$=0.76\pm{}0.30$, $0.78\pm{}0.26$;
    \item Random Forest: Accuracy$=0.83\pm{}0.21$, $0.84\pm{}0.21$;
    \item CNN (0.15$V_{A}$): Accuracy$=0.88\pm{}0.18$, $0.86\pm{}0.19$;
    \item CNN (0.10$V_{A}$): Accuracy$=0.85\pm{}0.23$, $0.87\pm{}0.20$;
    \item Always negative prediction: Accuracy$=0.79\pm{}0.30$.
\end{itemize}

In addition, we present the accuracy scores for always negative forecasts in the last line.
One can notice that Random Forest and k-nearest Neighbors are the two VDF moment-based classification models that slightly outperform the baseline always-negative forecast, with the Random Forest demonstrating a higher accuracy.
What is also interesting is that the deep learning-based CNN models outperform Random Forest, oppositely to the situation one has observed for the random train-validation splits of the data. This is also evident from Figures~\ref{figure:acc_allmodels}~and~\ref{figure:acc_allmodels_ps}: the CNN-based models rarely demonstrate lower accuracy scores with respect to the feature-based classification models, including Random Forest. All these results are pending the accurate assessment of the impact of the CNN training procedure described in Section~\ref{sec:MLmodel}. One last item to notice in the `unseen' simulation run tests is that the corresponding accuracy scores (highest of 0.88$\pm{}$0.18 for the CNN-based models) are significantly lower with respect to the best-performing Random Forest classifier on the random train-test split (0.96$\pm$0.01).

\section{Discussion and Conclusions}\label{sec:discussion}

In this study, we have utilized the dataset produced by hybrid-PIC simulations of the solar wind plasma with conditions representing observations by PSP near perihelia, and employed various ML / DL  techniques for the identification of ion-kinetic instabilities in the modeled SW plasma cases. 
Both feature-based classification models (kNN, SVM RBF, MLP, and Random Forest) operating on VDF statistical moments, and image-based convolutional neural networks (CNNs) operating on the VDFs directly have been tested. 
Our primary conclusions are the following:
\begin{itemize}
    \item The Random Forest classifier outperforms other classification models on the randomized train-validation split datasets (see Table~\ref{table:results_thres0001} and Figure~\ref{figure:scores}). 
    The corresponding accuracy is 0.96$\pm$0.01 and TSS score is 0.89$\pm$0.03 for the hyperparameters optimized with respect to the accuracy score.
    \item The empirical instability thresholds for the anisotropies and perpendicular magnetic energies of $>0.0005\Omega_{p}$ and $>0.001\Omega_{p}$ lead to the highest TSS scores of 0.90$\pm$0.02 and 0.89$\pm$0.01 (for models optimized for TSS, see Table~\ref{table:results_thresvar}) and manageable class-imbalance ratios of 915:681 and 1178:418. 
    Therefore, these thresholds lead to the `optimal' hybrid-PIC simulation run dataset for instability detection in the sense of performance of the ML models.
    \item For the tested performance on the `unseen' simulation runs (see Figure~\ref{figure:acc_allmodels}), the CNN-based model with the VDF velocity space resolution of $0.15V_{A}$ demonstrates the highest accuracy of $=0.88\pm{}0.18$, compared to the best-performing feature-based classifier Random Forest's accuracy of $=0.83\pm{}0.21$. 
    The averaged performances on the `unseen' runs are significantly weaker than on the random train-validation splits (accuracy of 0.96$\pm$0.01).
    \item The addition of the $E_{\perp}$ power spectrum that represents ion-scale wave activity as an input for the ML models does lead to the same or improved performance in all models (compare Tables~\ref{table:results_thres0001}~and~\ref{table:results_thres0001_ps}, Figures~\ref{figure:scores}~and~\ref{figure:scores_ps}), demonstrating that it could be beneficial to include the wave activity information in the instability analysis.
\end{itemize}

The fact that the Random Forest is the only model with prediction scores positively deviating with significance from all other models (see Tables~\ref{table:results_thres0001}~and~\ref{table:results_thres0001_ps} for details) for the random train-validation split of the data has several important implications. 
First, one can state with some confidence that the Random Forest model not only memorizes the data but also learns some patterns related to instability conditions from it. 
To remind us, for the random train-validation split, one can have the two snapshots from the same simulation neighboring in time designated into the train and test dataset, correspondingly. 
This can potentially generate an artificial correlation between datasets.
This adverse effect was recognized, for example, for the solar flare forecasting problem \citep{Ahmadzadeh2021ApJS..254...23A} and is called temporal coherence. 
However, we observe that Random Forest delivers significantly more accurate predictions with respect to the distance-based classifiers, such as kNN and SVM RBF (which would typically be assumed to perform well in the presence of temporal coherence). 
While there is definitely a correlation between the neighboring time moments of the considered simulation runs (which we cannot avoid in principle due to the limited number of simulations available), there is some evidence that Random Forest actually acts beyond the memorization of the various states.

\begin{figure}[htbp]
\centering
\begin{minipage}{0.49\linewidth}
    \centering
    \includegraphics[width=\linewidth]{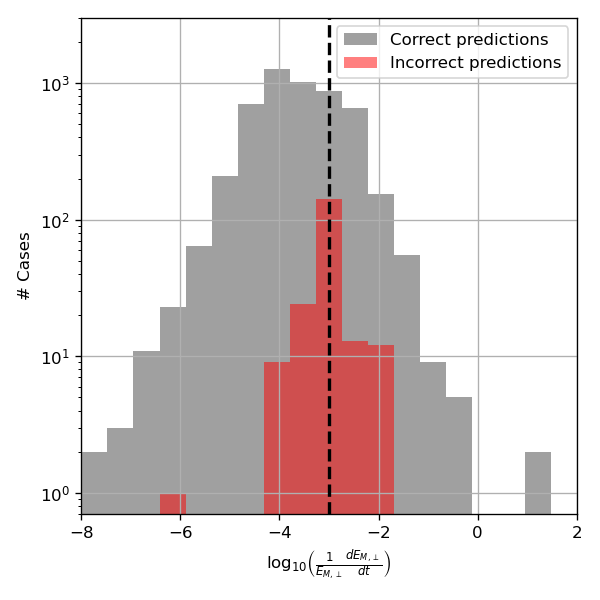}
    (a) 
\end{minipage}%
\begin{minipage}{0.49\linewidth}
    \centering
    \includegraphics[width=\linewidth]{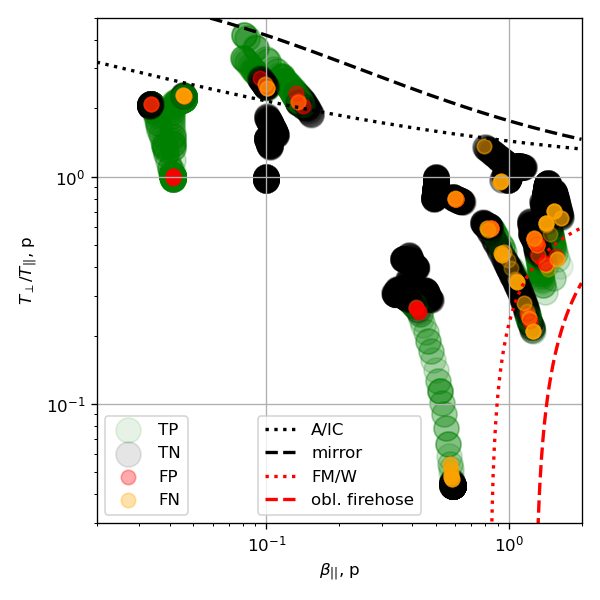}
    (b) 
\end{minipage}
\begin{minipage}{0.75\linewidth}
    \centering
    \includegraphics[width=1.00\linewidth]{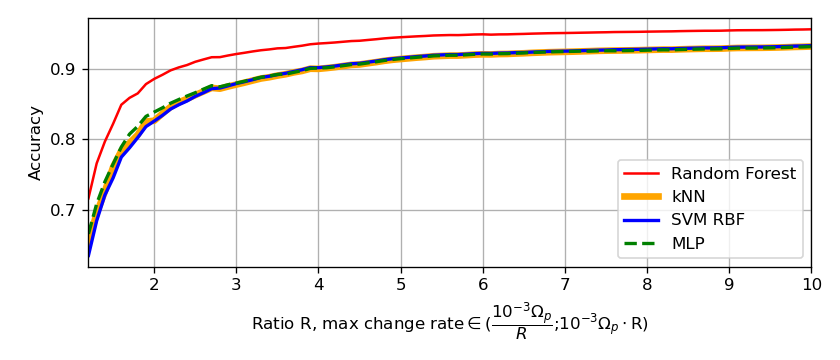}
    (c) 
\end{minipage}
\caption{Visualization of prediction outcomes across 10 validation datasets for the Random Forest classifier with the utilization of $E_{\perp}$ dynamic power spectrum. 
Panel (a) presents the rates of the magnetic energy change for the correct predictions and incorrect predictions. 
The black dashed line represents the empirical instability threshold used in this work.
Panel (b) visualizes the results in the $T_{\perp}/T_{\parallel}, p$ and $\beta_{\parallel}, p$ space. 
True positive (TP) predictions are shown in green, true negatives (TN) in black, false positives (FP) in red, and false negatives (FN) in orange.
Panel (c) demonstrated the accuracy of the feature-based models depending on the closeness of the change rate of the anisotropies or the perpendicular magnetic energy (whichever is greater) to the empirical instability threshold of 10$^{-3}\Omega{}_{p}$.
The dashed and dotted black and red lines in panel (b) represent the theoretical thresholds for the anisotropy-driven ion-kinetic instabilities following \citet{Verscharen2016collisionless} for the growth rate of $\gamma{}_{m}=10^{-3}\Omega{}_{p}$: ion-cyclotron (black dotter), mirror (black dashed), fast magnetosonic / whistler (red dotted), and oblique firehose (red dashed)
}
\label{figure:parameterspace}
\end{figure}

To attempt to understand the cases when the predictions were done incorrectly, we visualize the prediction outcomes across 10 validation datasets for the Random Forest classifier with the utilization of $E_{\perp}$ dynamic power spectrum in Figure~\ref{figure:parameterspace}. 
Panel (a) represents the perpendicular magnetic energy change rate. 
As one can see, the predictions are done incorrectly for the cases when the magnetic energy change rate was very close to the instability threshold, $\left\lvert{}\dfrac{1}{W_{M,\perp}}\dfrac{dW_{M,\perp}}{dt}\sim{}\right\rvert{}0.001\Omega{}_{p}$. 
Panel (b) represents the visualization of the prediction outcomes in the $T_{\perp}/T_{\parallel}, p$ and $\beta_{\parallel}, p$ space. 
One can see that the points form `ridges' in this figure, which corresponds to the evolution of individual simulation runs. 
One can also spot that in many cases the red and orange points (FP and FN, or incorrect predictions) appear in transitions from the green and black points (TP and TN, or correct predictions), again indicating that the performance of the ML model is mostly incorrect near the decision boundary. 
The behavior of the classifiers near the decision boundary is expected to be more uncertain, and some methodologies take that into account \citep{Ha2021} during the training process. 
We note here that other classifiers, such as SVM, demonstrate similar behavior in general, potentially indicating that Random Forest's better performance occurs because of more robust behavior near the decision boundary.
Figure~\ref{figure:parameterspace}c illustrates the dependence of the accuracy score on the closeness of the samples for which it is computed to the empirical instability threshold of $0.001\Omega{}_{p}$.
It is evident that the Random Forest continues to behave significantly better for the samples closer to the decision boundary with respect to the other three feature-based classification models tested.
Moreover, one can also notice that the absolute differences between the accuracy scores of Random Forest and other models increase closer to the decision boundary.

On the opposite side, the superior performance of Random Forest may point to the preference towards ensemble-based machine learning algorithms for the instability identification problem. 
We note here that the SAVIC ML model mentioned previously \citep{Martinovic2023ApJ...952...14M} utilizes XGBoost as a classifier, which is another ensemble method (boosting-type with respect to bagging-type Random Forest), demonstrating a very low number of false positive and false negative results and a superior accuracy of $\sim$0.96-0.99. 
A good performance of the ensemble-based learners was recognized in many other binary classification types of problems in application to space physics \citep[see, for example,][]{Ali2024ApJS..270...15A,OKeefe2024-RHclassifier}. 
It is interesting to note that the Random Forest method has weaker performance with respect to the CNN-based models for the experiment with `unseen' simulation runs (tests illustrated in Figure~\ref{figure:acc_allmodels}). 
This could indicate the potential of DL to capture the sophisticated non-linear relations in the VDFs leading to instabilities. 
We also note here that the CNN learning process involved weight decay regularization which we found necessary during the experiments in order to avoid overfitting. 
Potentially, this leads to preventing memorization and not-so-promising behavior for the random train-test splits, but good generalization capabilities and, correspondingly, good behavior on `unseen' runs. 
Further investigations are required to understand the exact reasons for such behaviors. 
However, it is definite that the VDF-based CNN models would be of interest in comparison with the classic feature-based models for kinetic instability detection for real-life scenario performance.

Lastly, the performance of all classifiers has dropped for the `unseen' run test with respect to the randomized train-validation split. 
As mentioned in Section~\ref{sec:results}, this could either indicate that the parameter space was extremely sparse and that the models have difficulty generalizing on the limited dataset, or that the temporal coherence takes place for the random train-validation splits. 
As discussed earlier and given the dimensionality of the initialization conditions for hybrid-PIC simulations, it does not seem that the similarity between the neighboring time moments is the only factor captured. 
The data sparsity is likely the issue here as well. 
This suggests that augmentation of the hybrid-PIC simulations' dataset is required; however, the simulations are computationally expensive, and some strategy is needed with respect to how to optimally sample the parameters space for new runs. 
The results presented in Figures~\ref{figure:acc_allmodels}~and~\ref{figure:acc_allmodels_ps} can provide hints for such strategies. 
For example, for some of the simulation runs the accuracy was lower than 50\% (random chance accuracy) for at least three models. These are the runs \#3, \#7, \#8, \#9, \#16, \#20, \#30.

\begin{figure}[htbp]
\centering
\begin{minipage}{0.49\linewidth}
    \centering
    \includegraphics[width=\linewidth]{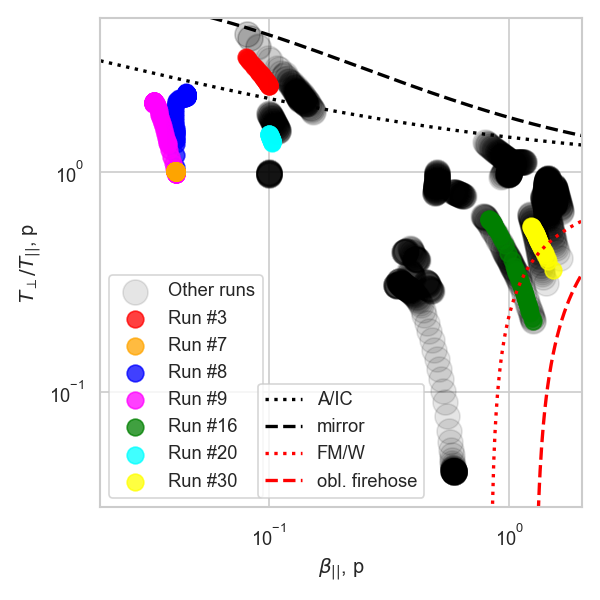}
    (a) 
\end{minipage}%
\begin{minipage}{0.49\linewidth}
    \centering
    \includegraphics[width=\linewidth]{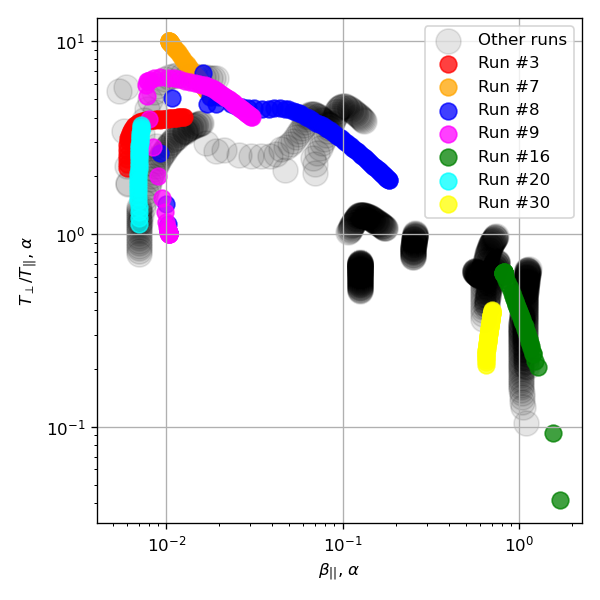}
    (b) 
\end{minipage}
\begin{minipage}{0.49\linewidth}
    \centering
    \includegraphics[width=\linewidth]{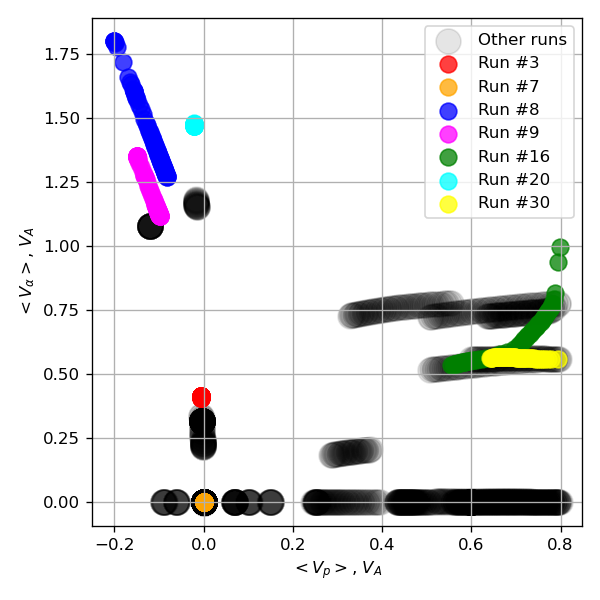}
    (c) 
\end{minipage}
\caption{Visualization of the evolution of the hybrid-PIC runs in the parameter spaces of (a) proton temperature anisotropy and $\beta{}_{||}$, (b) $\alpha$ temperature anisotropy and $\beta{}_{||}$, and (c) proton and $\alpha$ average velocities along the direction of the external magnetic field. Colored points represent the evolutions of runs for which the accuracy was lower than 50\% for at least three ML models. The black semi-transparent points represent the rest of the runs. The dashed and dotted black and red lines in panel (a) represent the theoretical thresholds for the anisotropy-driven ion-kinetic instabilities following \citet{Verscharen2016collisionless} for the growth rate of $\gamma{}_{m}=10^{-3}\Omega{}_{p}$: ion-cyclotron (black dotter), mirror (black dashed), fast magnetosonic / whistler (red dotted), and oblique firehose (red dashed).}
\label{figure:parameterspace_runs}
\end{figure}

The dynamics of these simulation runs in the selected components of the parameter space are visualized in Figure~\ref{figure:parameterspace_runs}. 
While it is generally difficult to derive strong conclusions given the high-dimensional parameter space where the runs are initialized and evolve, one can notice some patterns. 
For example, the runs \#7, \#8, and \#9 have proton temperature anisotropies $>$1 and, at the same time, low $\beta{}_{\parallel} < 5$$\times{}10^{-2}$, all away from the populations of the majority of well-predicted runs, as visible in Figure~\ref{figure:parameterspace_runs}a. 
The runs \#3, \#16, \#20, and \#30 are relatively close to well-predicted runs in the same parameter space but start to differ from them in Figures~\ref{figure:parameterspace_runs}b (runs \#3 and \#30) and~\ref{figure:parameterspace_runs}c (runs \#3, \#8, \#20, and partially \#16). 
While it is hard to provide quantitative conclusions on this point, one can assume that the initialization of new simulation runs with the parameters in the vicinity of the aforementioned runs should, in principle, bring more VDF samples in the parameter space where poor performance was observed. 
Therefore, such an `unseen' run analysis could represent a viable strategy for targeted enhancement of the pool of simulation runs.

\section*{Acknowledgements}
This work has been supported by the NSF SHINE grant AGS-2300961 and NASA HTMS grant 80NSSC24K0724. VMS also thanks the NSF FDSS grant 1936361. Computational resources supporting this work were provided by the NASA High-End Computing (HEC) Program through the NASA Advanced Supercomputing (NAS) Division at Ames Research Center. PM acknowledges the partial support by NASA HGIO grant 80NSSC23K0419 and the NSF SHINE grant 2401162.

\bibliography{references}{}
\bibliographystyle{aasjournal}

\end{document}